\documentclass{vgtc}                          

\ifpdf
  \pdfoutput=1\relax                   
  \usepackage{graphicx}                
  \DeclareGraphicsExtensions{.pdf,.png,.jpg,.jpeg} 
\else
  \usepackage{graphicx}                
  \DeclareGraphicsExtensions{.eps}     
\fi%

\graphicspath{{figures/}{pictures/}{images/}{./}} 

\usepackage{microtype}                 
\PassOptionsToPackage{warn}{textcomp}  
\usepackage{textcomp}                  
\usepackage{mathptmx}                  
\usepackage{times}                     
\usepackage{cite}                      
\usepackage{tabu}                      
\usepackage{booktabs}                  

\usepackage{setspace}
\usepackage[table]{xcolor}
\usepackage{hyperref, url}
\hypersetup{colorlinks = true, linkcolor = cyan, filecolor = cyan, urlcolor = cyan, citecolor = cyan}
\usepackage{amssymb, amsmath}
\usepackage{enumerate, caption, tikz}
\newcolumntype{M}[1]{>{\centering\arraybackslash}m{#1}} 
\usepackage{graphicx}
\usepackage{array,multirow,graphicx}
\usepackage{arydshln}
\usepackage{makecell}
\usepackage[latin1]{inputenc}
\usetikzlibrary{shapes,arrows}
\usepackage[edges]{forest} 
\usepackage{forest} 
\usepackage{subcaption} 

\onlineid{0}

\vgtccategory{Research}

\title{Visualizing Income Distribution in the United States}

\author{Sang T. Truong\thanks{e-mail: sttruong@cs.stanford.edu}\\
     \scriptsize Stanford University
     \and Humberto Barreto\thanks{e-mail: hbarreto@depauw.edu}\\
        \scriptsize DePauw University
     }

\teaser{\centering
    \includegraphics[width= 0.32\textwidth]{figures/RPPERHHINCOME_1976.jpg}
    \includegraphics[width= 0.32\textwidth]{figures/RPPERHHINCOME_1998.jpg}
    \includegraphics[width= 0.32\textwidth]{figures/RPPERHHINCOME_2019.jpg}
    \caption{The first perspective shows a significant development of household income with adjustment $H$ from 1976 to 2019 in the U.S. Each slide along the percentile axis (the shortest one) represents a state. The median national household income distance (the benchmark) determines each state's color in 2019. The red or blue colors represent a state below or above the benchmark in 1976, respectively. We take advantage of the colors to give viewers an overall sense of position-changing patterns over time.}\label{hhat}}

\abstract{The distribution of household income is a central concern of modern economic policy due to its strong influence on life quality. Yet, non-expert audiences are unaware of the relationship between these two factors. To effectively communicate the effect of income inequality on the quality of life and among the strata, we have designed a novel technique for visualizing income distribution and inequality over time by using the U.S. household income microdata from the Current Population Survey. The result is a striking dynamic animation of income distribution over time, drawing public attention and further investigating economic inequality~\footnote{Detailed implementation of this project is available at https://github.com/sangttruong/incomevis}.}

\CCScatlist{
  \CCScatTwelve{Human-centered computing}{Visualization}{Visualization application domains}{Information visualization}; }

\begin{document}
\maketitle
\section{Introduction}
Although income has always affected the life standard for a long time, the Americans may not understand this impact properly. Indeed, higher-income correlates to longer life expectancy~\cite{chetty2016}. The advent of modern economic growth in the last few decades has further elevated concerns about income inequality~\cite{heathcote2010, gudrais2014}. The Americans have lived through a marked uptick in the share of income going to the richest among us~\cite{Piketty2014, SaezZucam2019, attanasio2012, reynolds2007, dunhaupt2014, gudrais2014}. Unfortunately, most of us are unaware of the tremendous change in income distribution within the U.S.~\cite{norton2011}. When surveyed about perceived income differences by race and ethnicity, the evidence indicated that the magnitude of these misperceptions was substantial. The respondents estimate that for every \$100 in wealth held by a White family, a Black family has \$90, when, in reality, the Black family only has \$10; misperceptions about the Latin-White wealth gap are just as significant as Black-White~\cite[p. 917]{kraus2019}.

\begin{figure}[ht]\centering
    \includegraphics[width = 0.65\columnwidth]{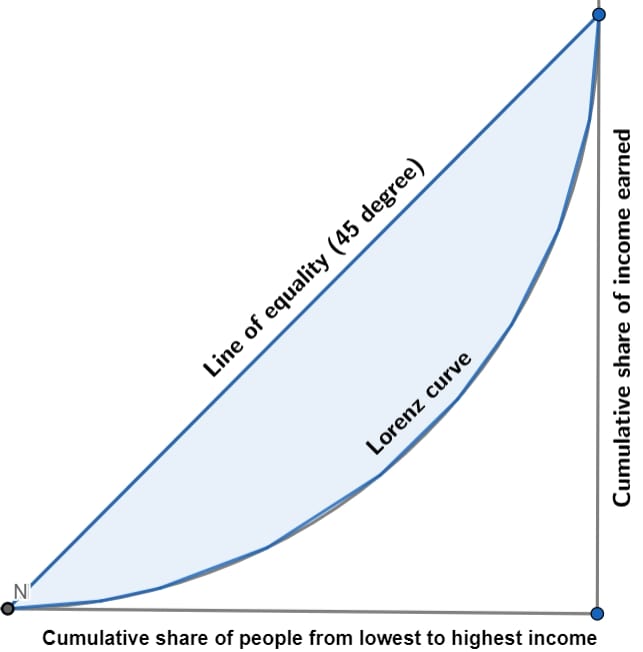}
    \caption{The Lorrenz curve~\cite{Aaberge2001}.}\label{lorenz}
\end{figure}

Regardless of its significance, the quantity of income distribution is still ambiguous to non-experts. The most common way to convey inequality in income distribution is via the Gini coefficient (G)~\cite{Dorfman1979, hartmann2017}. For a discrete income distribution with $n$ entities, the Gini coefficient is expressed as:
\begin{equation}
    G = \frac{\sum_{i=1}^n{\sum_{j=1}^n{|x_i - x_j|}}}{2\sum_{i=1}^n{\sum_{j=1}^n{x_j}}} = \frac{\sum_{i=1}^n{\sum_{j=1}^n{|x_i - x_j|}}}{2n^2\bar{x}}
    \label{ginicoeff}
\end{equation}
where $x_i$ is the income of entity $i$ and $\bar{x}$ is the average income. Since $G$ depends on the area between Lorenz curve and a diagonal line (Figure~\ref{lorenz}), its value is between 0 and 1. $G=0$ indicates a perfectly egalitarian distribution where everyone has the same income, and $G=1$ indicates one person receives all of the income. While these two extremes are relatively easy for non-expert to understand, the values between the interval of $G$ are more difficult to interpret because it adopts a nonlinear scale. For example, $G = 0.5$ does not indicate an "average" degree of inequality but rather a high one. In addition, decreasing 0.1 from $G = 0.3$ is not the same if it occurs from 0.6. Not only does $G$ require some experience to interpret, but it also has a few undesirable properties. For example, various distributions of income give the same value of $G$ because these entire distributions fall into a single number. Due to Gini coefficients' complexity, nonprofessionals may struggle to understand the structure of income.

This paper proposes a visualization framework that demonstrates income distribution and inequality over time to audiences without background knowledge. Our visualization provides an instant and accurate snapshot of current inequality. Also, it yields historical context to compare present to past differences. Unlike the Gini coefficient, it does not require advanced statistical knowledge but instead presents the income inequality in a clear and  informative way.

\section{Related Works}
\begin{figure}[t]\centering
    \includegraphics[width = 0.75\columnwidth]{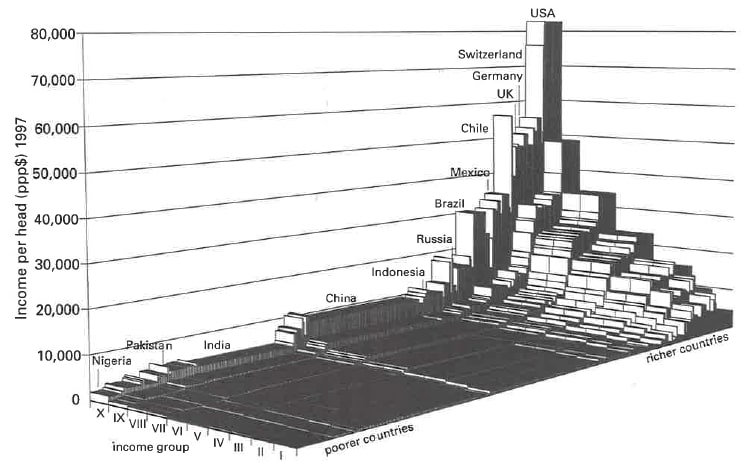}
    \caption{Global income distribution~\cite{sutcliffe2001}.}\label{sutcliffe}
\end{figure}

Visualization of global income inequality  (reproduced in Figure~\ref{sutcliffe}) shows the world distribution of income as a city~\cite{sutcliffe2001, coreecon2020}. Most prominently, in the far corner, the skyscraper represents the income of the wealthiest $10^{th}$ percent of the world population~\cite[p. 17]{sutcliffe2001}. Income inequality not only shows in different countries but also within a specific country. China is an example that suffers severely from income inequality mainly due to public policies~\cite{han2011}. The U.S. is another country experiences this problem.

Although income inequality in the U.S. is relatively popular among economists, only a few attempts to communicate this information to non-expert audiences. In 2013, Roser showed different income inequality computations not only in the U.S. but also in various countries~\cite{roser2013}. Unlike Roser, Lawrence focused more on analyzing income inequality within the U.S. but with different demographic aspect~\cite{lawrence2018}. Our analysis is the combination of Lawrence's work and Figure~\ref{sutcliffe} to illustrate the income inequality more precisely, simply, and clearly to non-expert audiences.

\section{Data Processing}
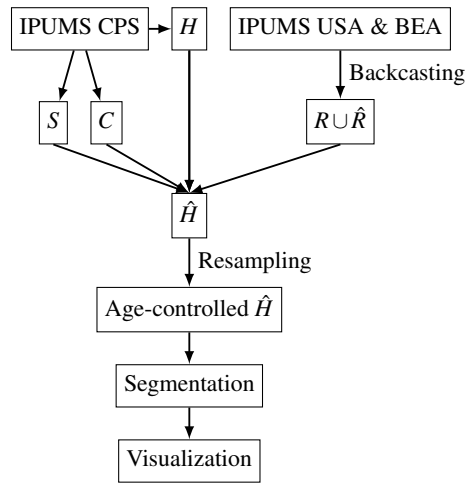
\begin{figure}[t]\centering
    \begin{forest}
        for tree={s sep=3mm}
        [, for tree={draw, edge={-latex, line width=0.25mm}}, phantom
        [IPUMS CPS, align=center, name = x
        [$S$, l=1.25cm, align=center, name = F2]
        [$C$, l=1.25cm, align=center, name = F3]]
        [$H$, align=center, name=m,
        [$\hat{H}$, l=2.5cm, align=center, name=down,
        [Age-controlled $\hat{H}$,  l=1.25cm, edge label={node[midway,right]{Resampling}}, align=center,
        [Segmentation,  l=1cm, align=center,
        [Visualization, l=1cm,  align=center,
        ]]]]]
        [IPUMS USA \& BEA, align=center
        [$R \cup \hat{R}$, l = 1.25cm, edge label={node[midway,right]{Backcasting}}, align=center, name=F6]]
        ]
        \draw[-latex, line width=0.25mm] (F2.south) -- (down.north);
        \draw[-latex, line width=0.25mm] (F3.south) -- (down.north);
        \draw[-latex, line width=0.25mm] (F6.south) -- (down.north);
        \draw[-latex, line width=0.25mm] (m.south) -- (down.north);
        \draw[-latex, line width=0.25mm] (x) -- (m);
    \end{forest}
    \caption{Data processing framework}\label{dataprocessing}
\end{figure}

Figure~\ref{dataprocessing} describes our data processing framework for U.S. data. Other datasets from different countries can reuse this framework for various economic variables as well (e.g.\ personal income or cumulative wealth).

\begin{table}[t]
    \resizebox{0.46\textwidth}{!}{
    \def\arraystretch{1.25}
    \begin{tabular}
        {| m{.075\textwidth} | m{.19\textwidth} | M{.06\textwidth} | M{.06\textwidth} | M{.06\textwidth} | M{.06\textwidth} | }\hline
        Variables & Description & $\mu$ & $\sigma$ & min & max \\\hline
        Household income ($H$) & Total nominal income of all household members during previous year & 52343.81 & 65094.54 & -37040 & 3299997 \\\hline
        Consumer price index (C) & Estimation of inflation base on 1999 price level & 1.18 & 0.55 & 0.65 & 2.92 \\\hline
        Regional price parity (R) & The differences in price levels across states and metropolitan areas for a given year and are expressed as a percentage of the overall national price level & 97.53 & 8.54 & 84.8 & 119.2 \\\hline
        Gross rent (r) & Gross monthly rental cost of the housing unit, including contract rent plus additional costs (e.g. utilities) & 841.98 & 201.19 & 512.0 & 1600.0 \\\hline
        Effective household size (S) & Squared root of number of household member & 1.58 & 0.45 & 1 & 5.10 \\\hline
        Household weight (w) & Household-level weight that should be used to generate statistics about households in March Annual Social and Economic (ASEC) Supplement data & 1,538.89 & 920.93 & 0.00 & 17957.53 \\\hline
    \end{tabular} }
    \caption{Summary statistics in the household levels. Each variable has a total of 2.82 million observations for the 1976-2019 period.}\label{summarystat}
\end{table}

\begin{figure}[hbt!]\centering
    \begin{subfigure}{.49\columnwidth}\centering
        \includegraphics[width = \columnwidth]{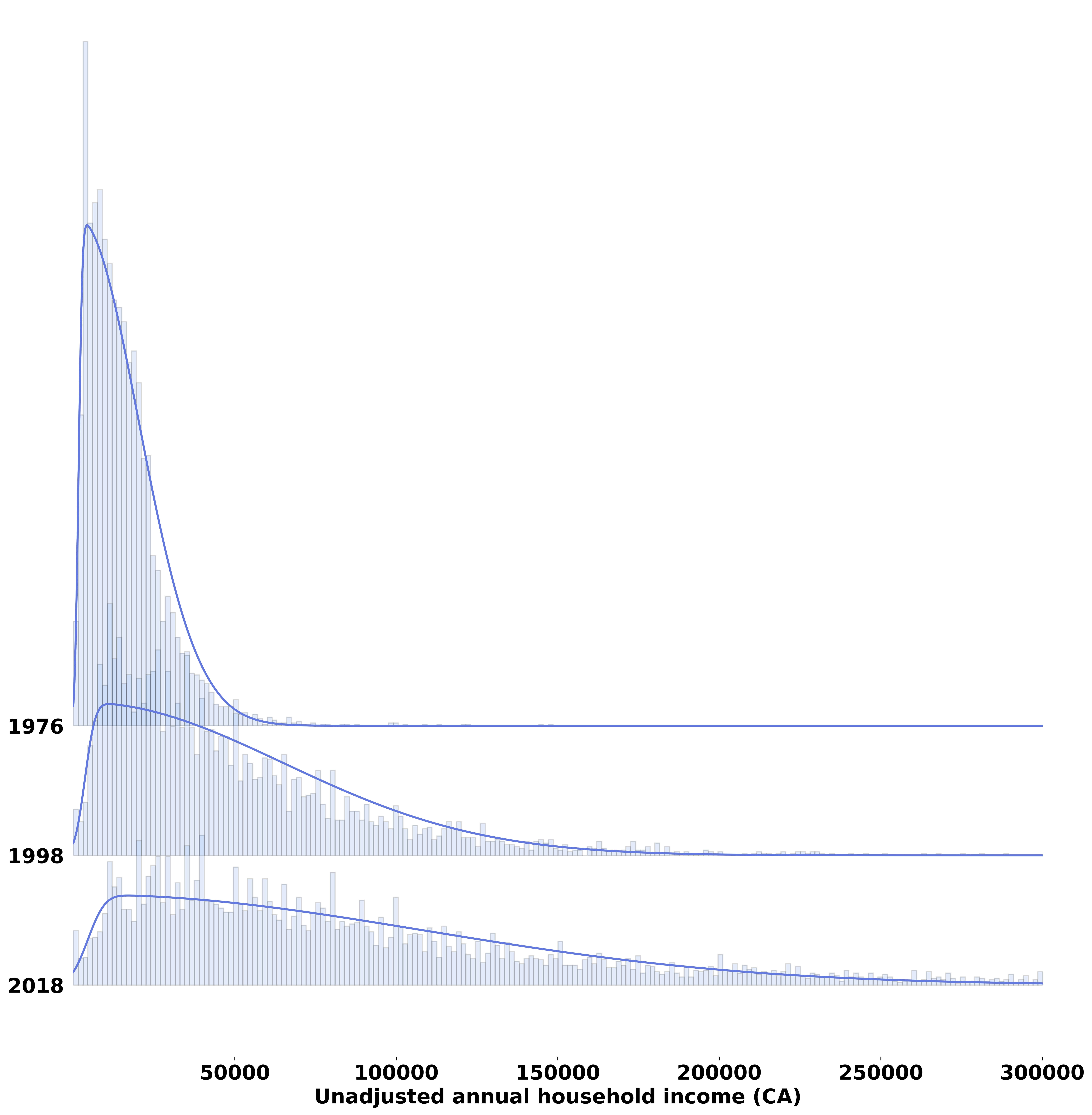}
        \caption{Distribution of $H$ in CA over time}\label{hhincome-6}
    \end{subfigure}
    \begin{subfigure}{.49\columnwidth}\centering
        \includegraphics[width = \columnwidth]{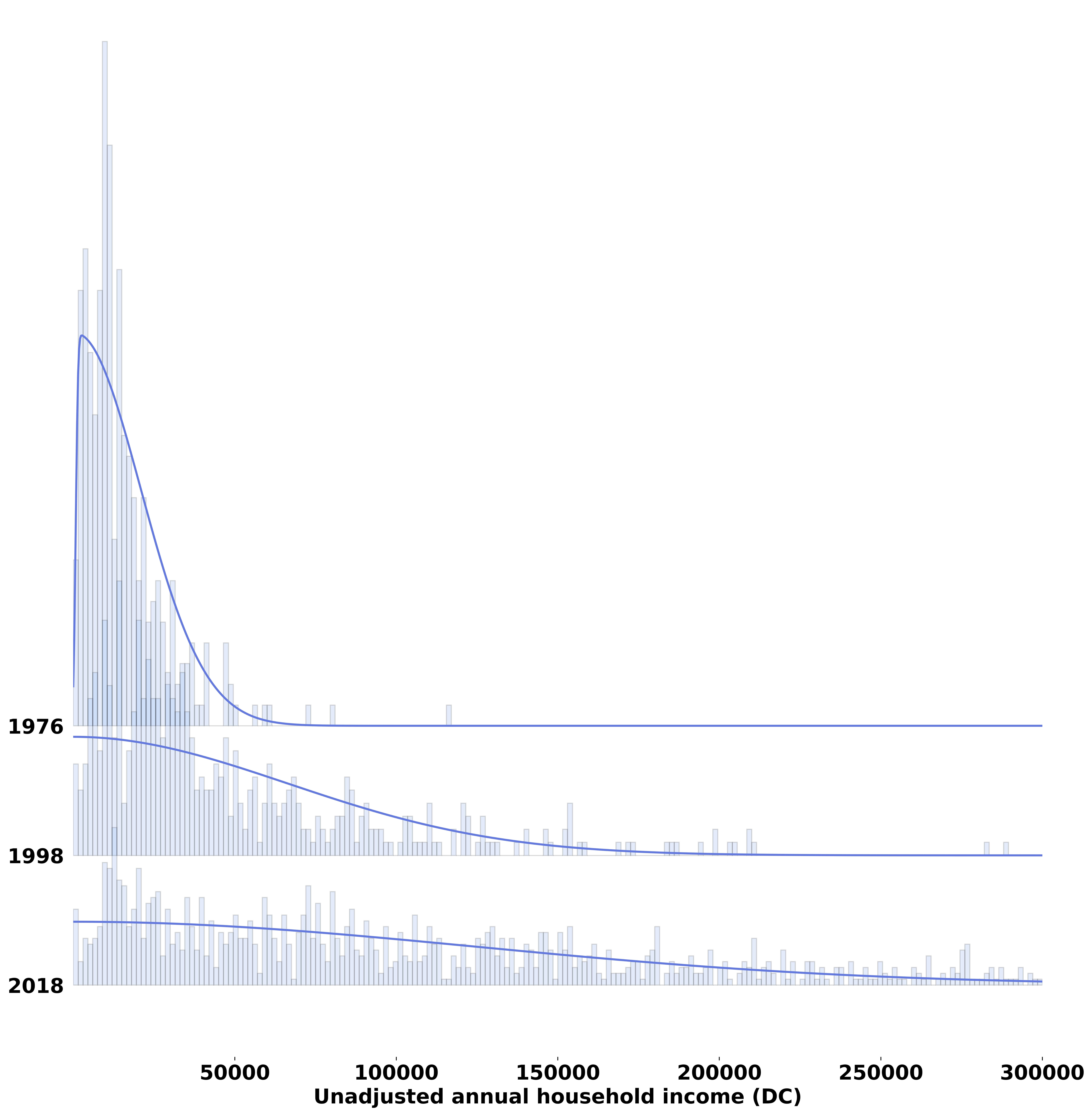}
        \caption{Distribution of $H$ in DC over time}\label{hhincome-11}
    \end{subfigure}
    \begin{subfigure}{.49\columnwidth}\centering
        \includegraphics[width = \columnwidth]{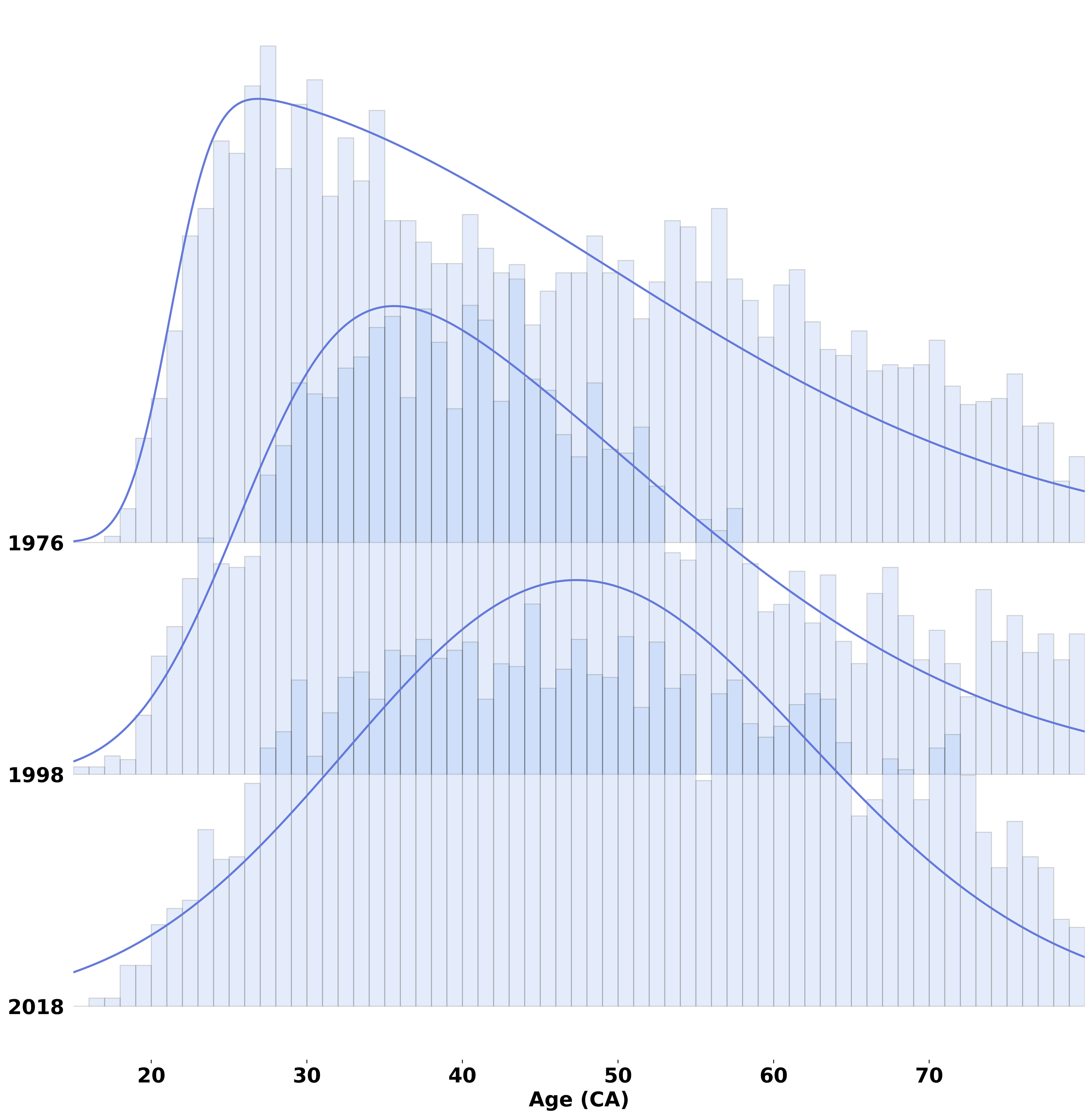}
        \caption{Distribution of age in CA over time}\label{age-6}
    \end{subfigure}
    \begin{subfigure}{.49\columnwidth}\centering
        \includegraphics[width = \columnwidth]{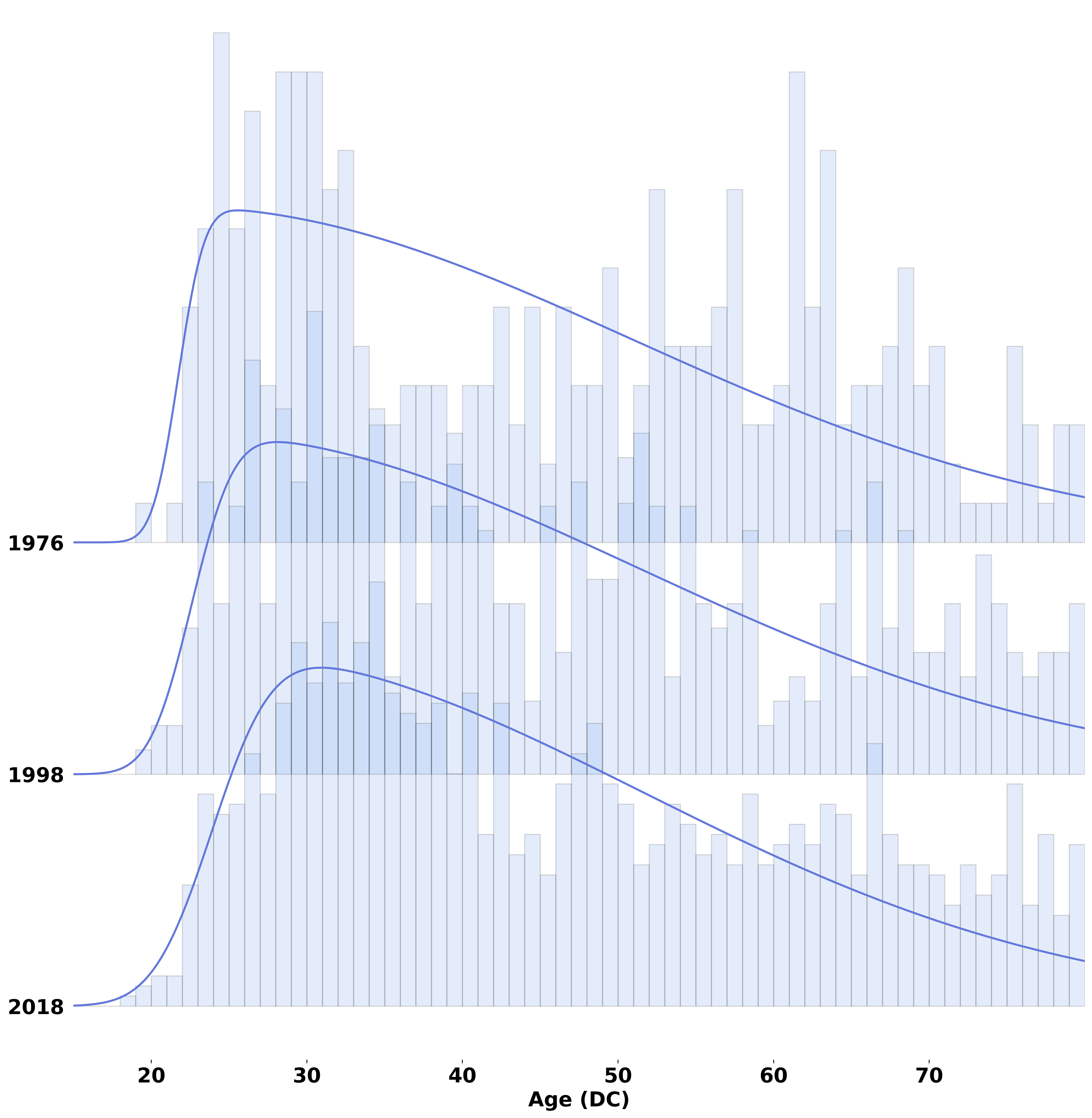}
        \caption{Distribution of age in DC over time}\label{age-11}
    \end{subfigure}
    \caption{The income distribution of both CA and DC has been increasingly right-skewed. The household size of both states is relatively stable over time. CA has a larger average household size than DC.}\label{hhsize-ca-dc}
\end{figure}

\subsection{The Data}

\begin{figure*}[t]
    \begin{subfigure}{.24\textwidth}\centering
        \includegraphics[width = \textwidth]{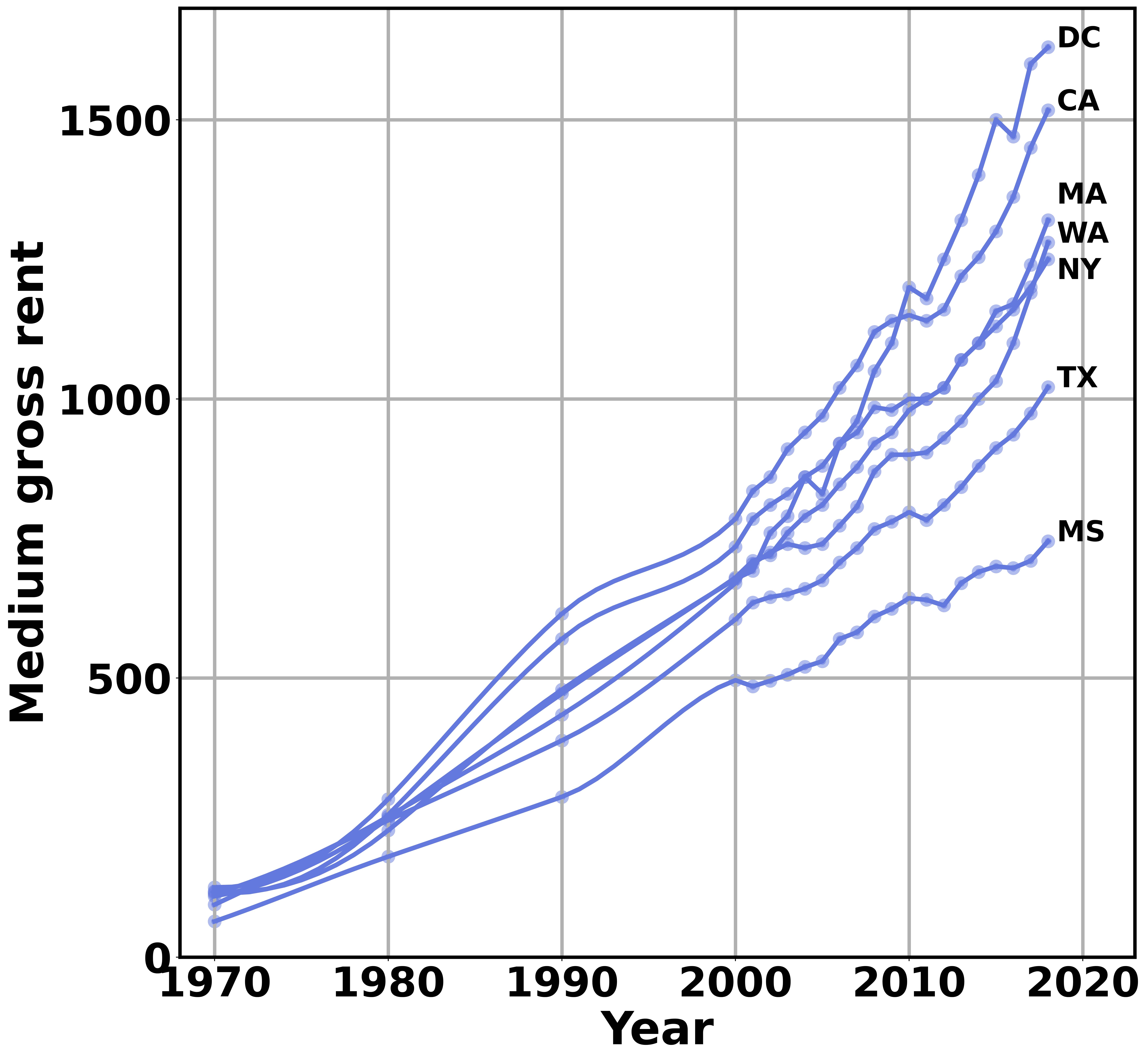}
        \caption{Gross rent ($r$ and $\hat{r}$)}\label{rent}
    \end{subfigure}
    \begin{subfigure}{.24\textwidth}\centering
        \includegraphics[width = \textwidth]{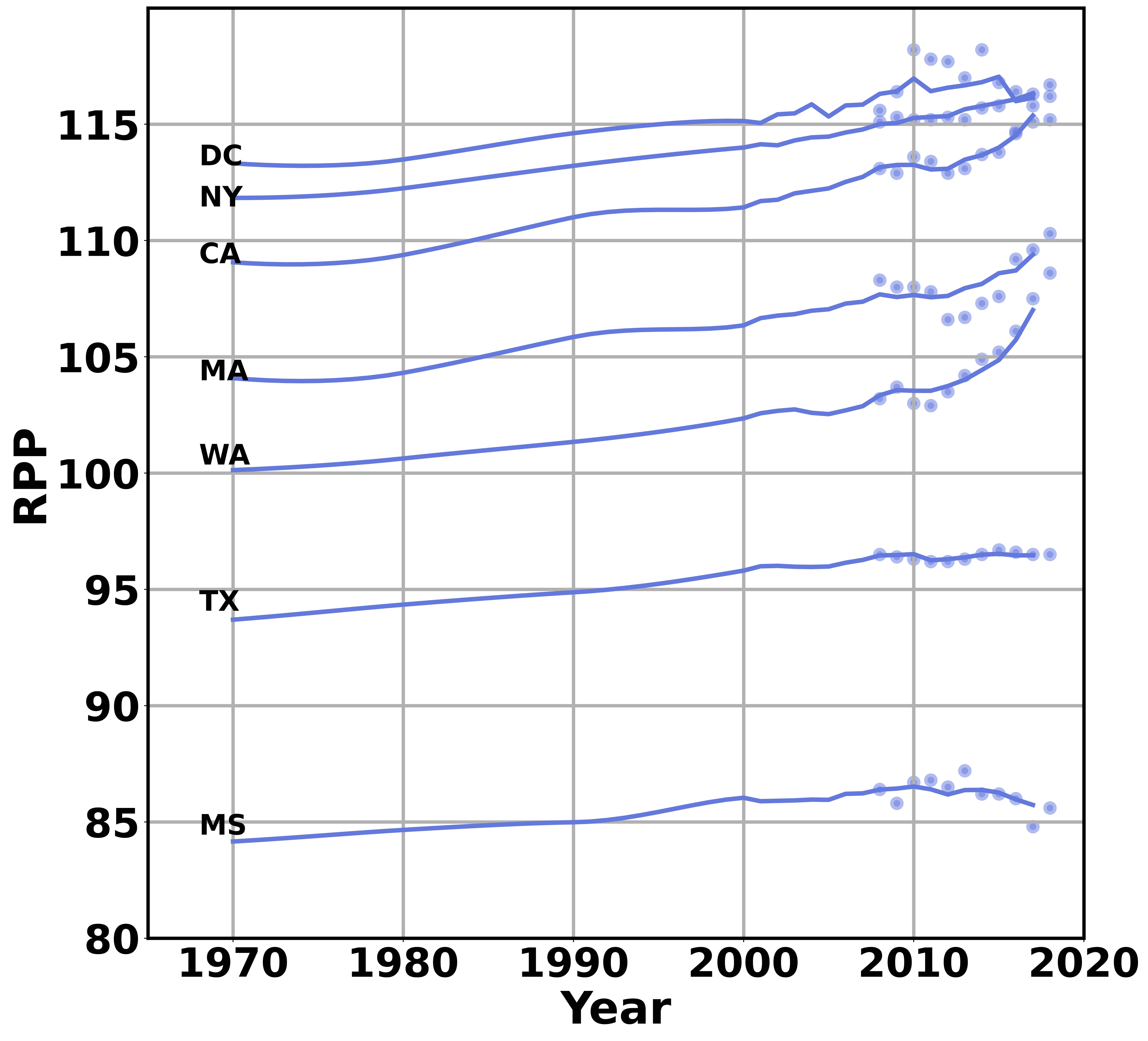}
        \caption{Regional price parity ($R$ and $\hat{R}$)}\label{rpp}
    \end{subfigure}
    \begin{subfigure}{.24\textwidth}\centering
        \includegraphics[width = \textwidth]{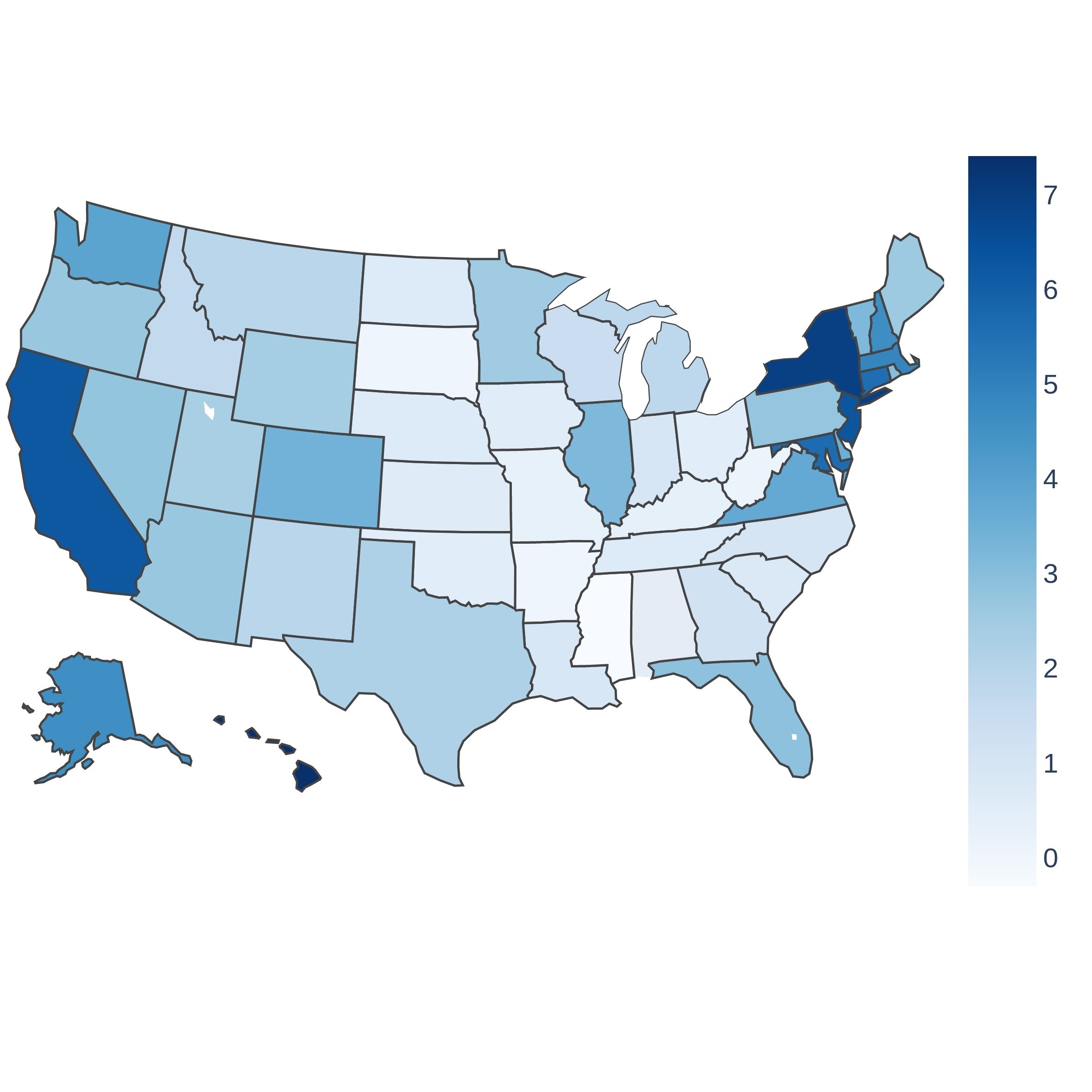}
        \caption{State fixed effect of regional price parity}\label{fe-coef}
    \end{subfigure}
    \begin{subfigure}{.24\textwidth}\centering
        \includegraphics[width = \textwidth]{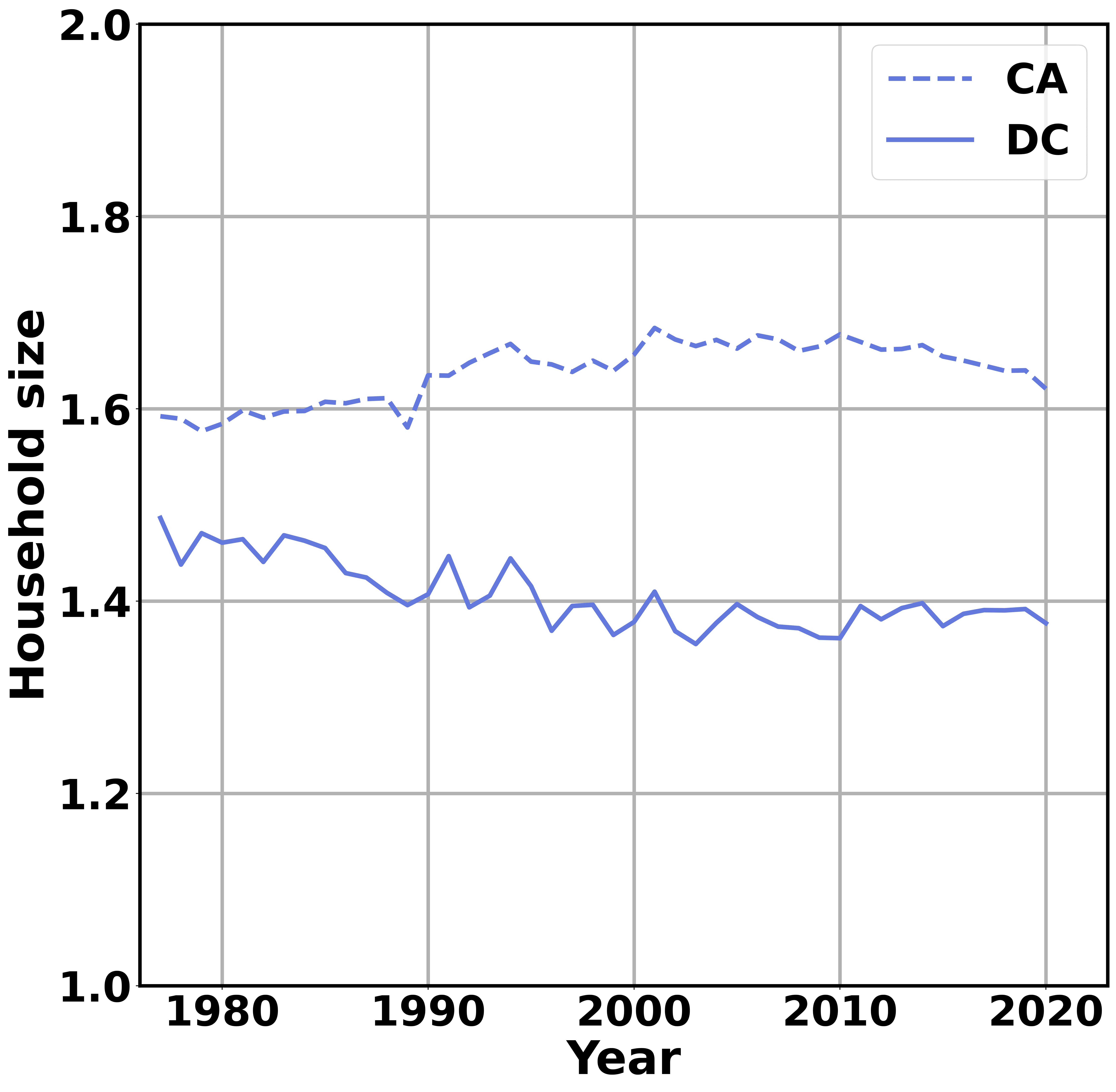}
        \caption{Average effective household size $S$}\label{hhsize-6-11}
    \end{subfigure}
    \caption{Estimation of regional price parity for each state from 1976 to 2019 applies model \ref{backcast}. Gross rent has a linear trend in the majority of states. The fixed effect coefficients in the backcasting model show that DC and CA are among the states with the highest living expense.}\label{rpp-backcast}
\end{figure*}

\begin{figure*}[hbt!]\centering
    \begin{subfigure}{.24\textwidth}\centering
        \includegraphics[width = \textwidth]{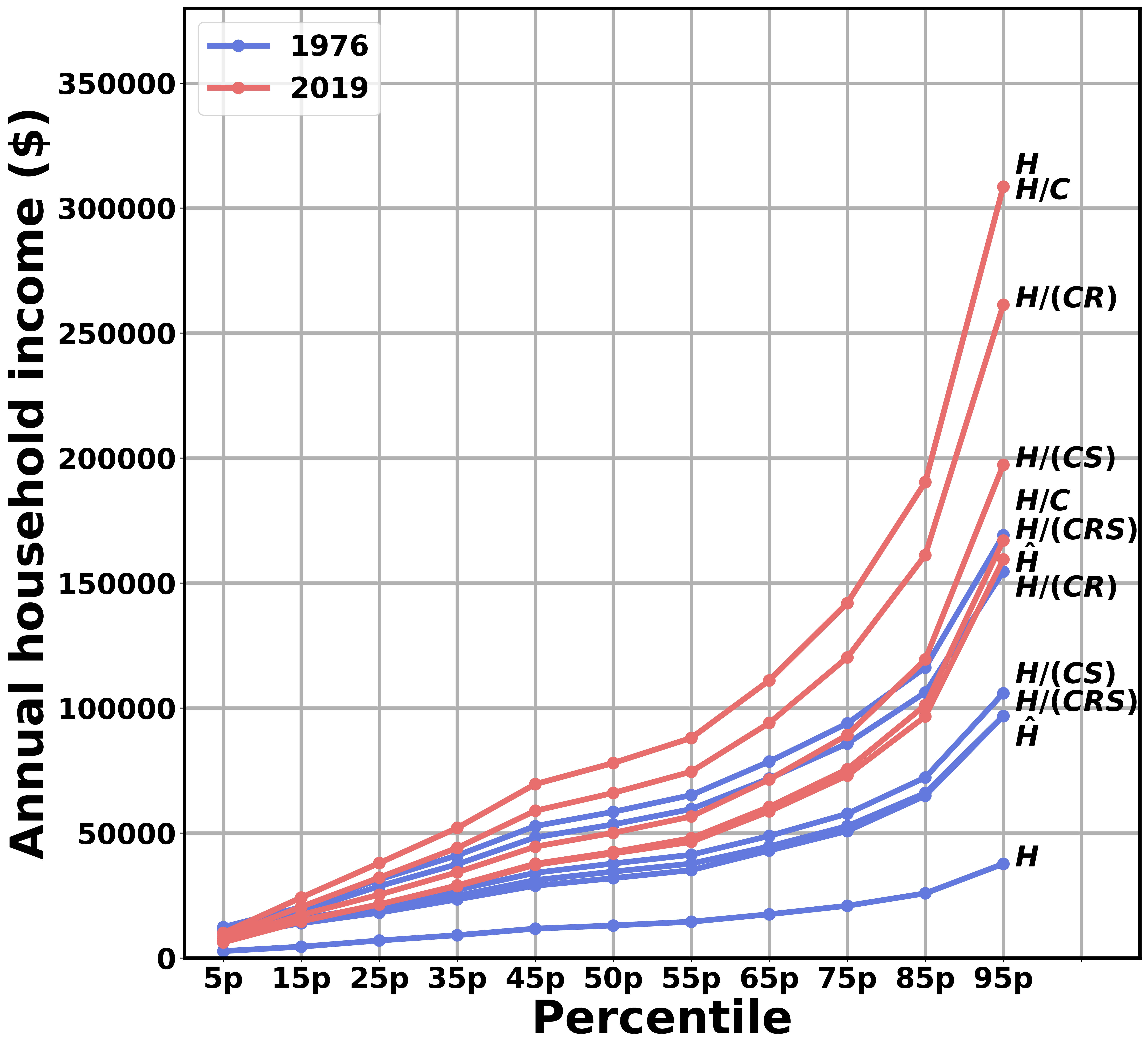}
        \caption{Various normalizers for CA}\label{CA_all_income}
    \end{subfigure}
    \begin{subfigure}{.24\textwidth}\centering
        \includegraphics[width = \textwidth]{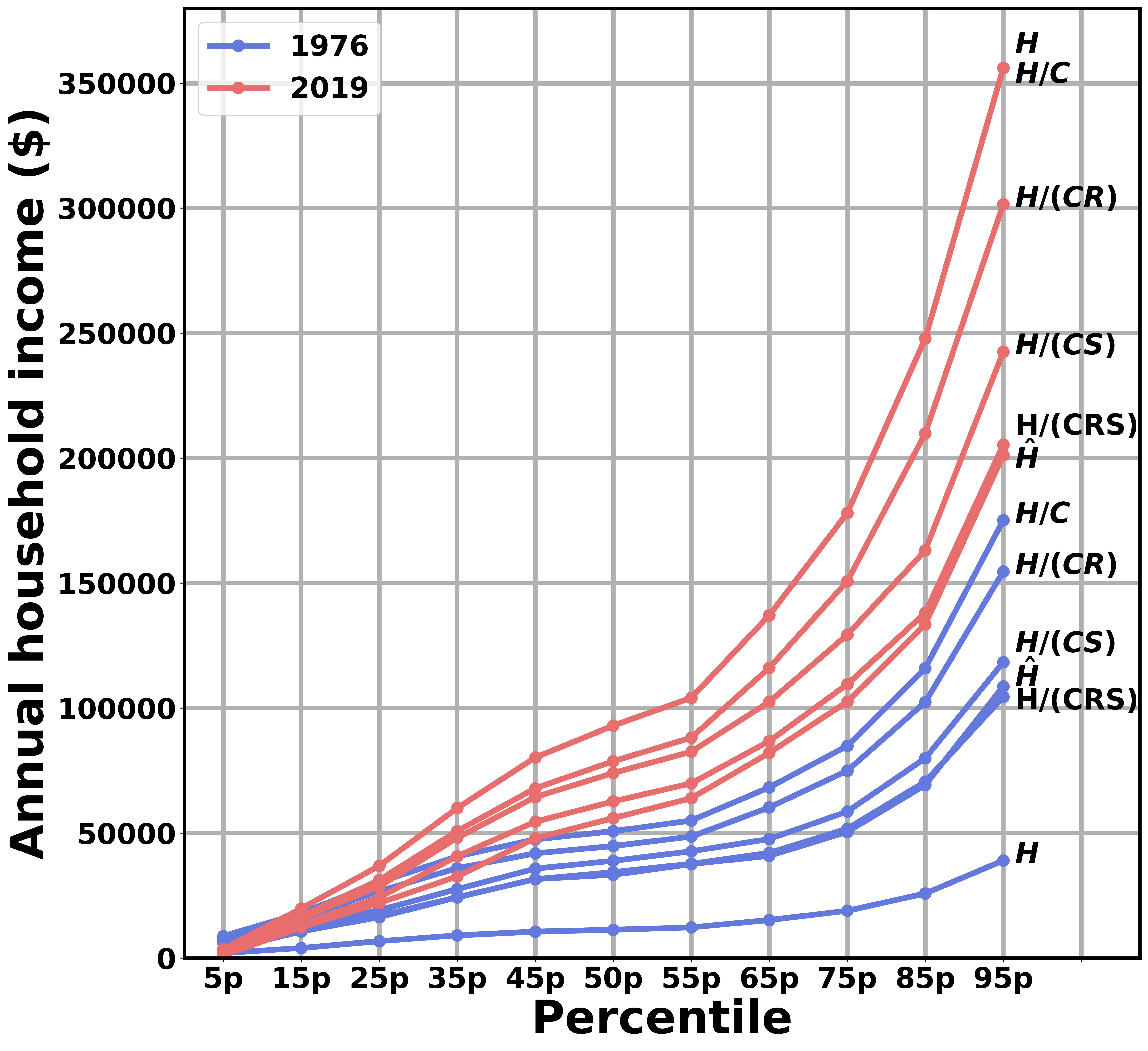}
        \caption{Various normalizers for CA}\label{DC_all_income}
    \end{subfigure}
    \begin{subfigure}{.24\textwidth}\centering
        \includegraphics[width = \textwidth]{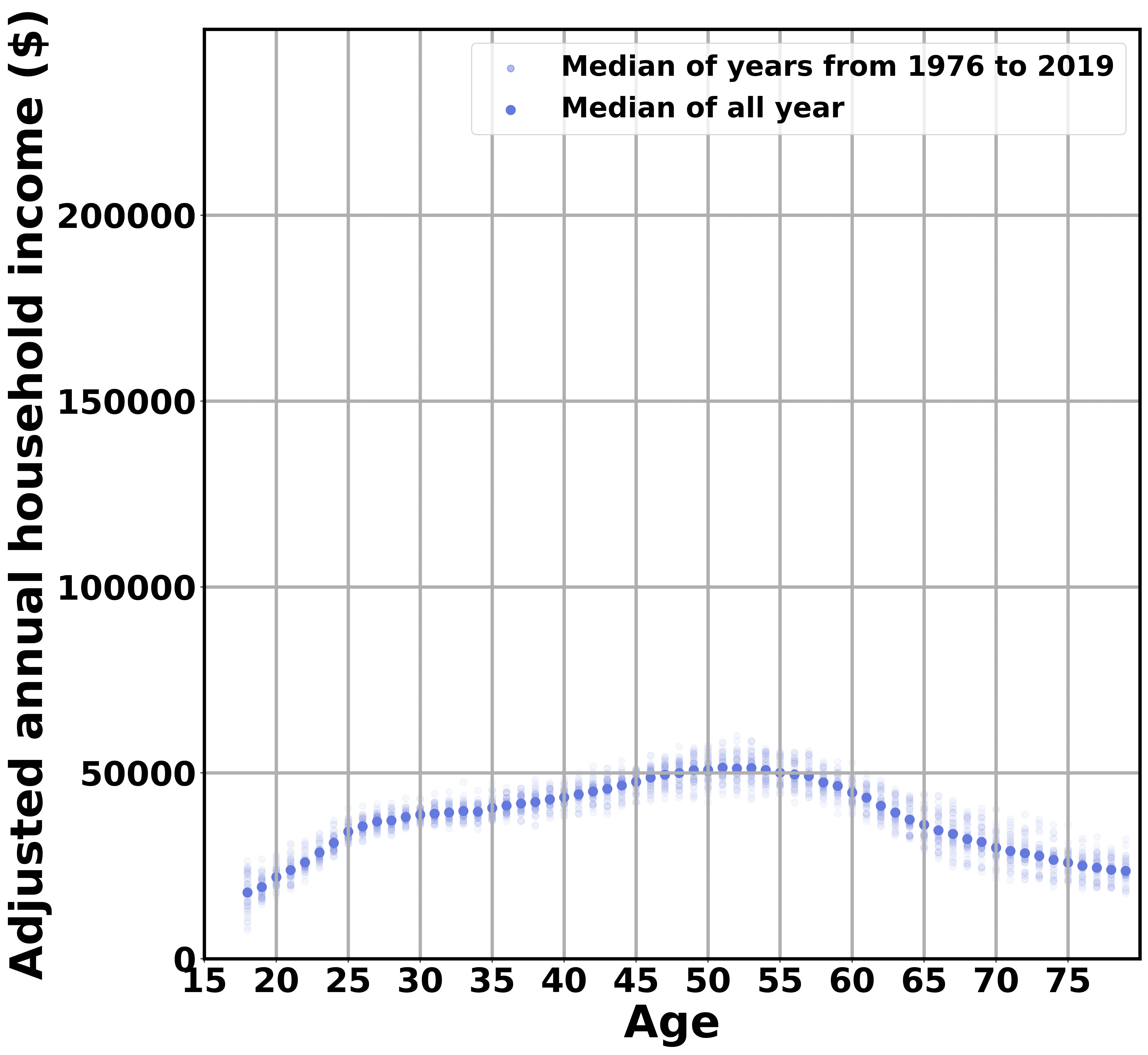}
        \caption{Median $\hat{H}$ for each age (1976-2019)}\label{all_income}
    \end{subfigure}
    \begin{subfigure}{.24\textwidth}\centering
        \includegraphics[width = \textwidth]{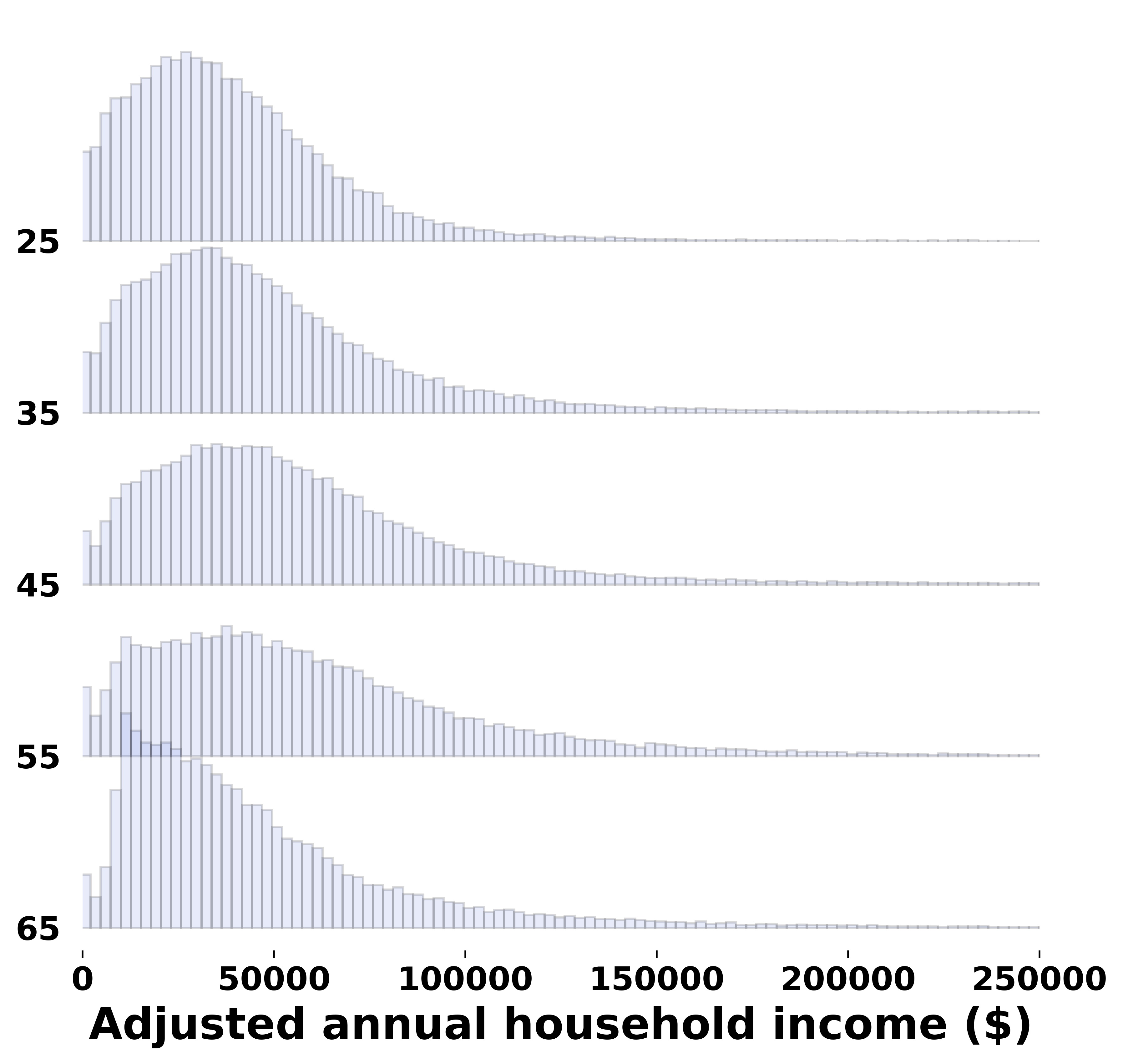}
        \caption{$\hat{H}$ distribution for each age (1976 - 2019)} \label{age-income-relationship}
    \end{subfigure}
    \caption{The age distributions differ across states. DC has a relatively younger population than CA. The age distribution of each state may not be stable over time. Age and income imply a strong relationship in the U.S.}\label{age-income}
\end{figure*}
Our data is collected from Integrated Public Use Microdata Series - Current Population Survey (IPUMS-CPS), Integrated Public Use Microdata Series - United States of America (IPUMS USA), and the Bureau of Economic Analysis (BEA), from 1977 to 2020~\cite{flood2020, bea2020, ruggles2020}. Since population survey data reported in year $t$ is for the previous year, $t-1$, our analysis is for the 1976-2019 period. We only extend the analysis back to 1976 because the geographic information is not reliable before this time. Table~\ref{summarystat} presents the summary statistics of the data. Information about regional price parity from BEA~\cite{bea2020} allows a meaningful comparison of income across different states, which is hard to generate without this dataset due to the cost of living in various regions (e.g.\ it is much more expensive to live in CA than to do so in AL). Although BEA database is the most complete version for regional price parity in the U.S. that we have access to, it is only available annually from 2008.

\subsection{Normalizers for Household Income}
We adjust for the geographic differences in prices to make better comparisons of income between states. For example, people in CA earn higher wages than in AL, but the cost of living in CA is more expensive than in AL. BEA offers regional price parities $R$ to enable the comparisons of buying power across the country~\cite{bea2020}. To retrieve the unavailable data of $R$ before 2008, we perform backcasting with the following model:
\begin{equation}
    R = \alpha + \beta_1 r + \beta_2 F(r) + \beta_3 F(R)+ \beta_{4,i} \textit{FE}_i + \epsilon
    \label{backcast}
\end{equation}
where $F(\cdot)$ is the forward-shift operator (i.e.\ lead operator); $FE_i$ is the binary fixed effect for every state $i$; and $r$ is the gross rent. We choose $r$ as the primary independent variable to predict $R$ because it is the main component of the cost of living within the U.S. Moreover, since $r$ does not have data annually before 2000, we interpolate to recover the missing data.

\begin{figure*}[hbt!]\centering
    \includegraphics[width= 0.32\textwidth]{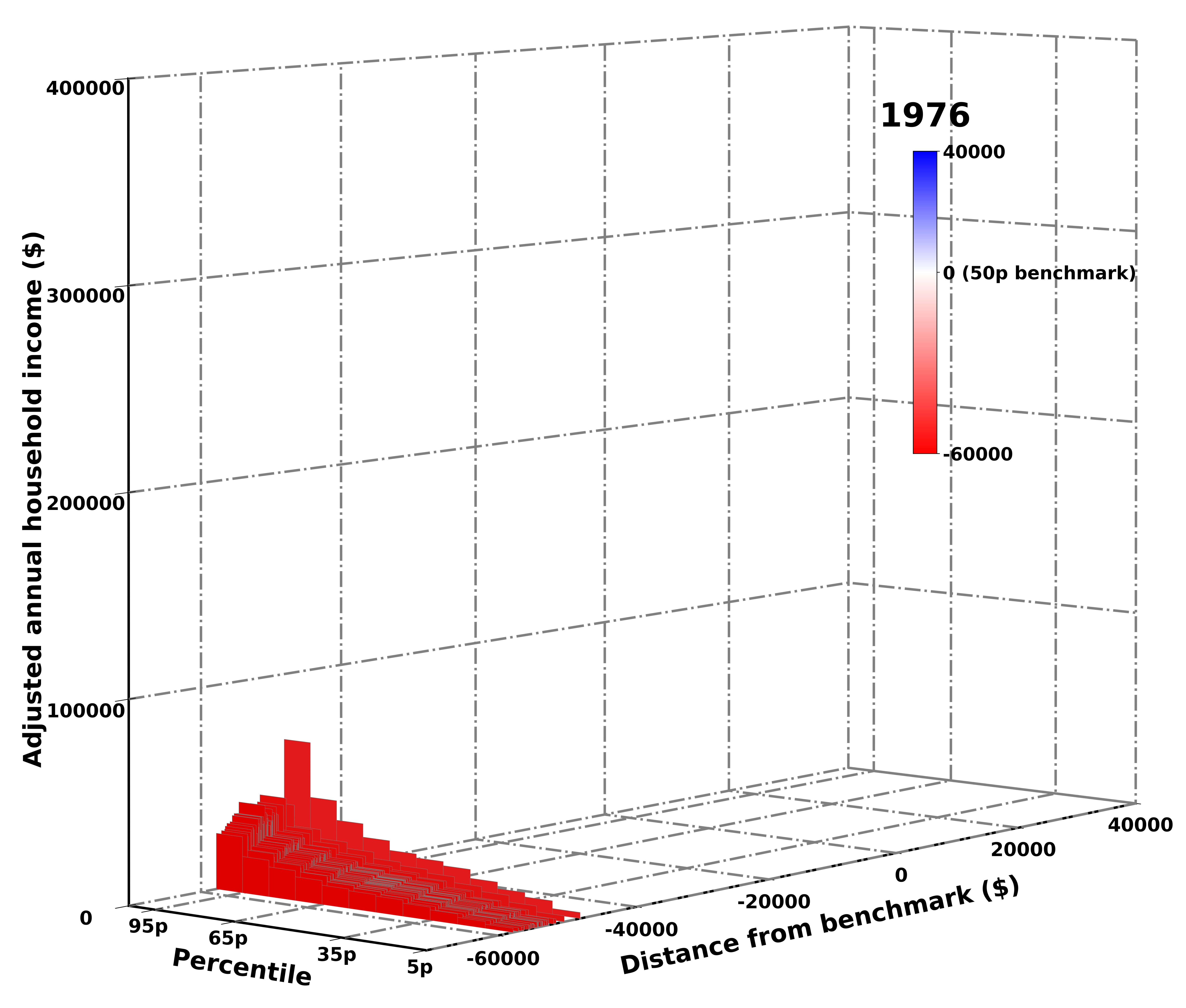}
    \includegraphics[width= 0.32\textwidth]{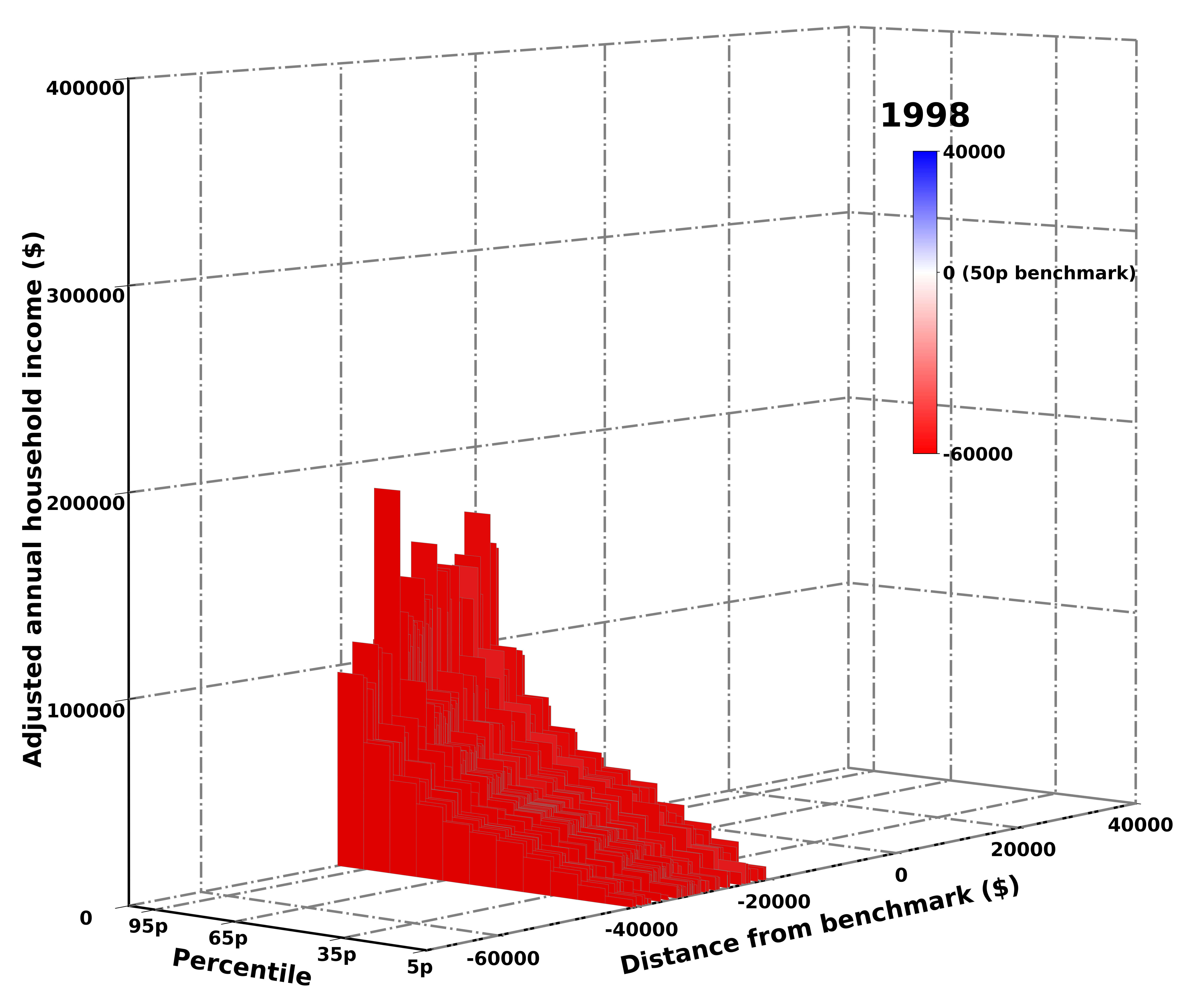}
    \includegraphics[width= 0.32\textwidth]{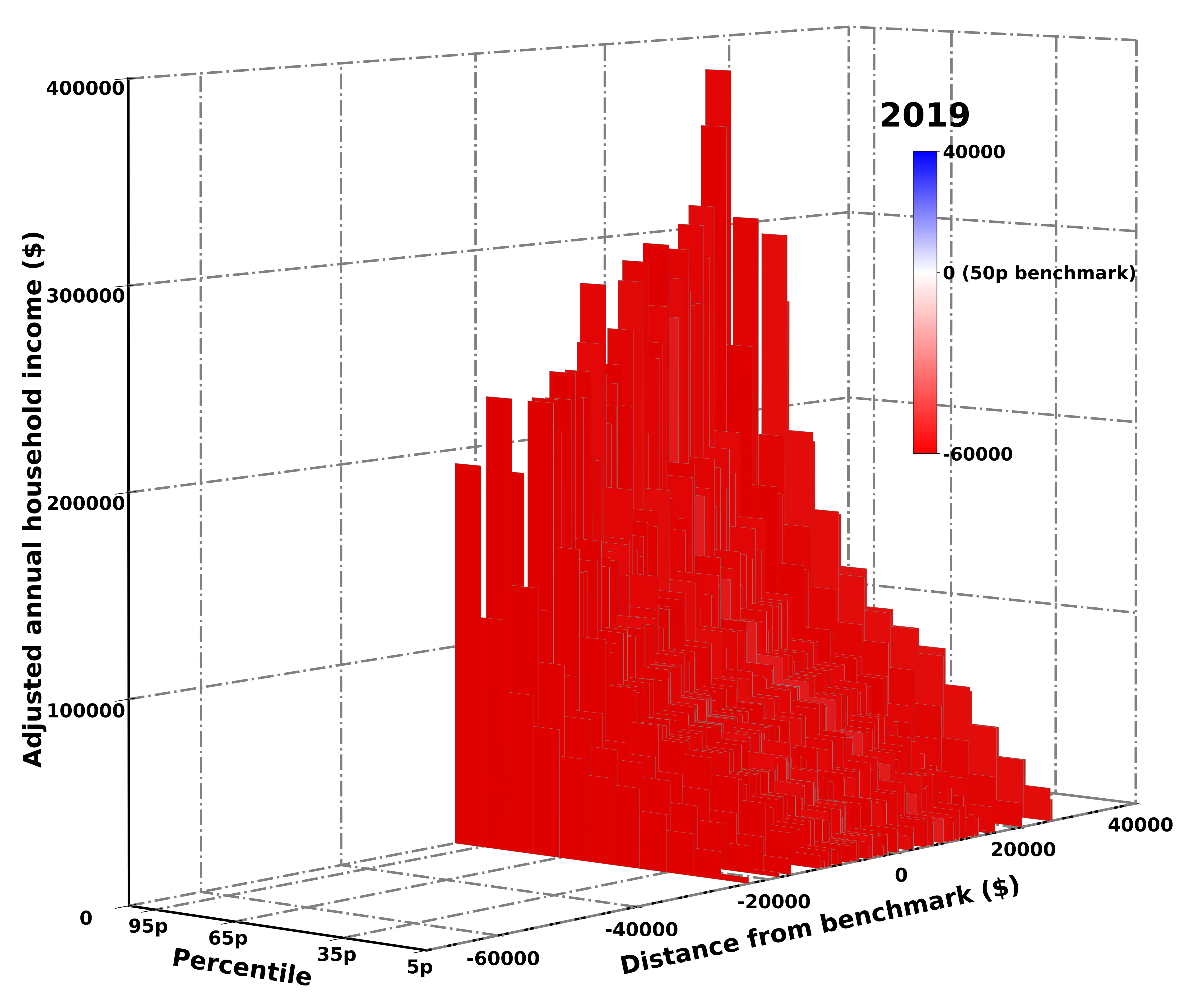}
    \caption{The second perspective demonstrates the development of household income without adjustment from 1976 to 2019 in the U.S. This movement illustrates an illusion of growth.}\label{h}
\end{figure*}
\begin{figure*}[hbt!]\centering
    \includegraphics[width= 0.32\textwidth]{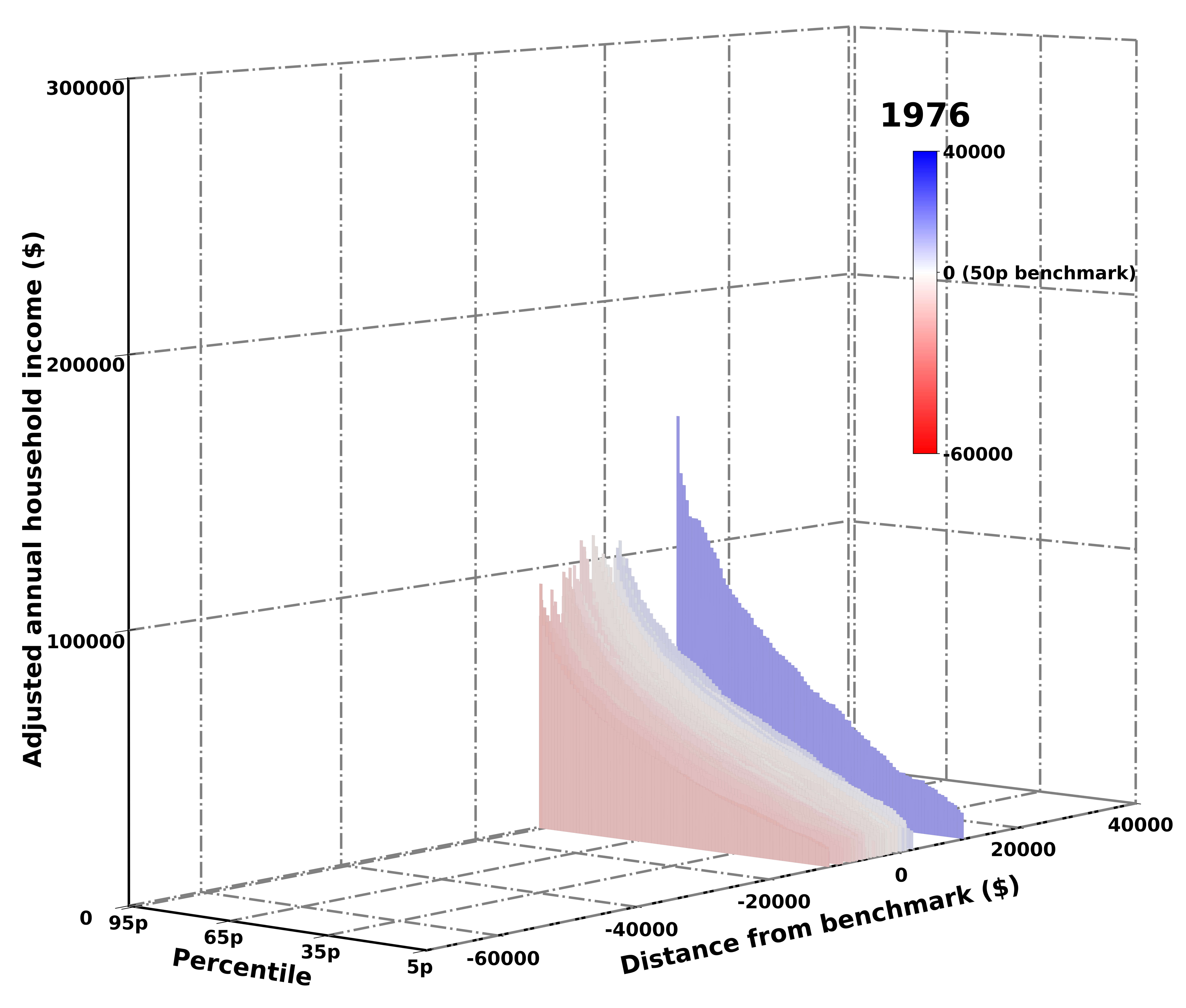}
    \includegraphics[width= 0.32\textwidth]{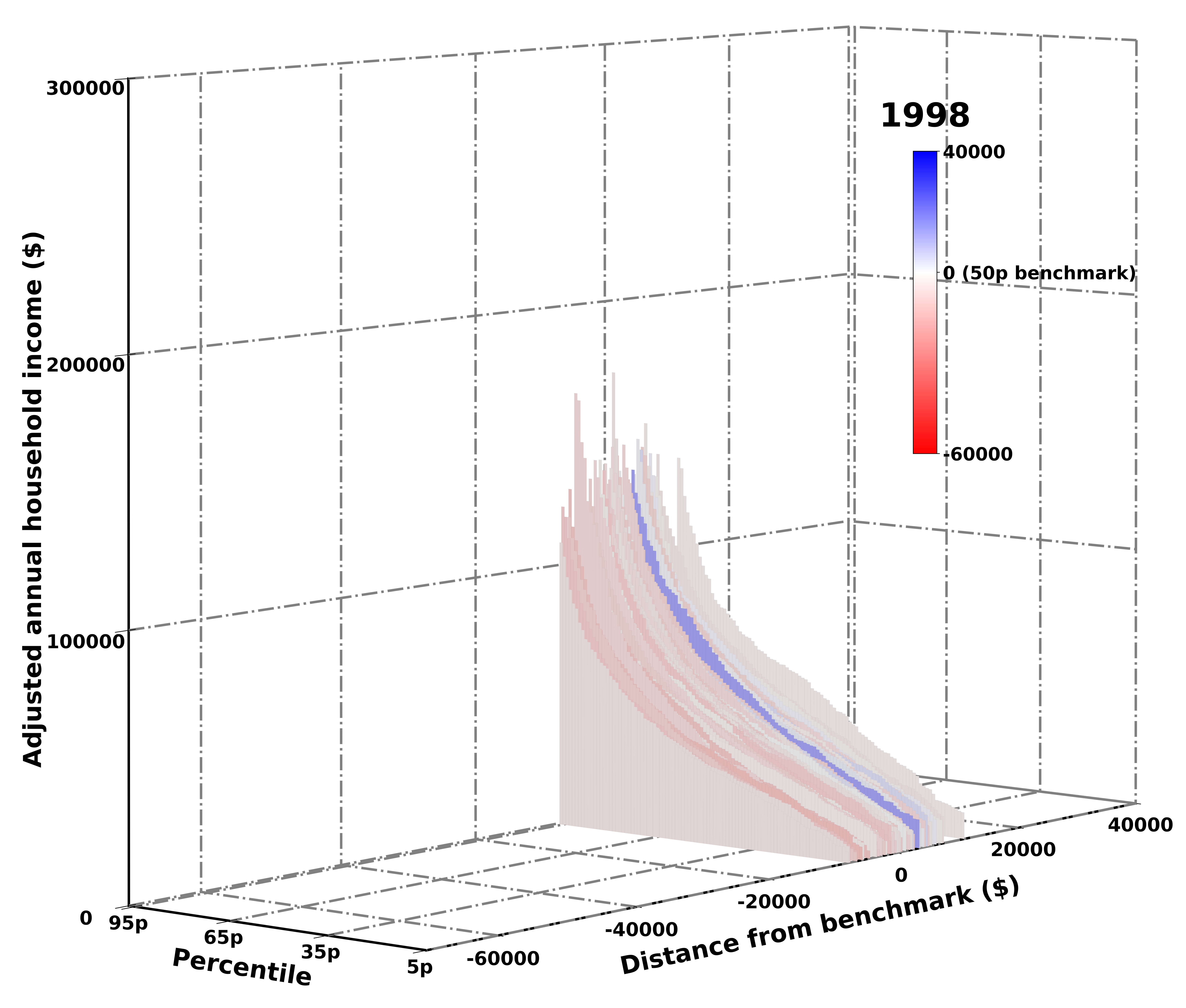}
    \includegraphics[width= 0.32\textwidth]{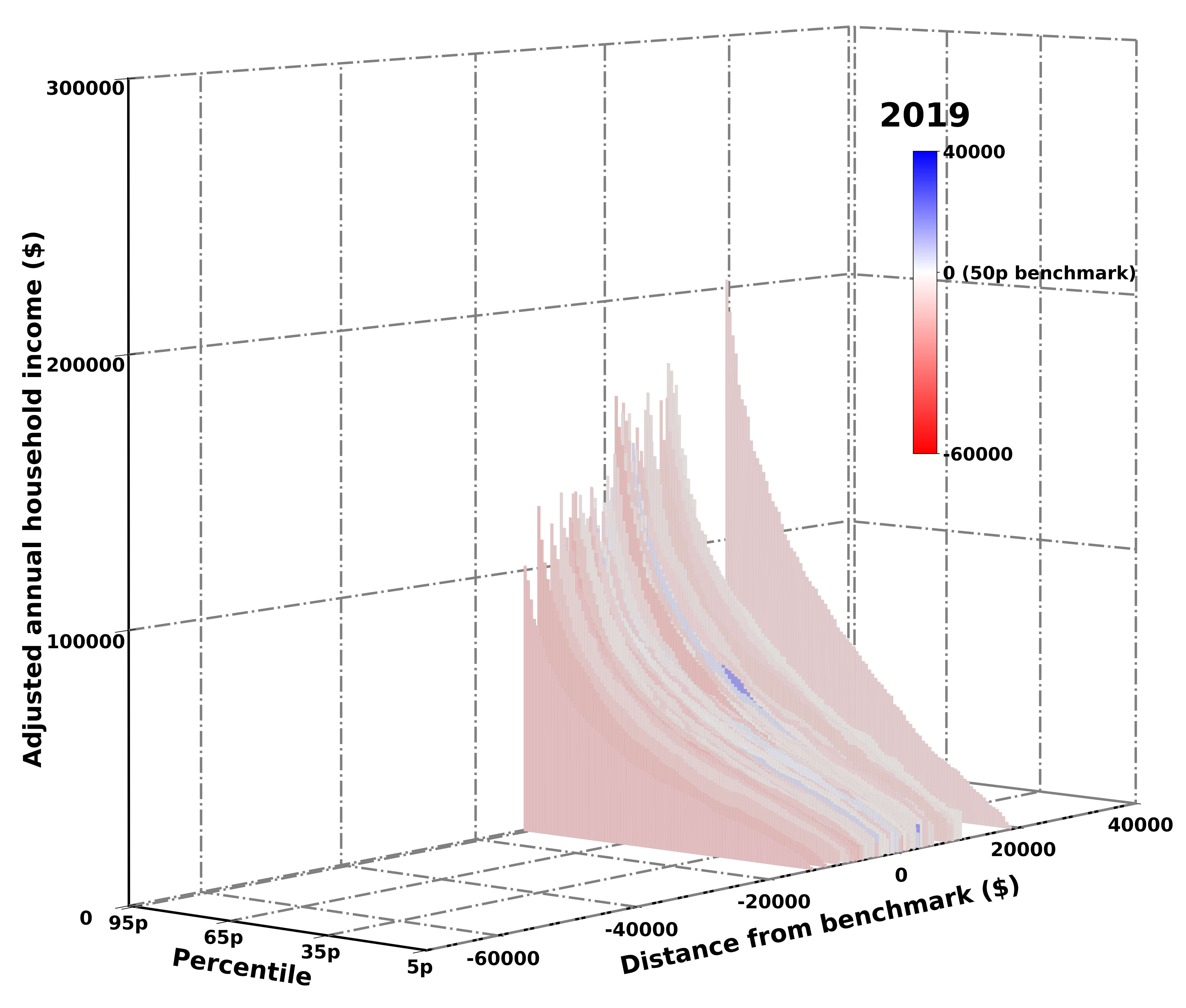}
    \caption{The third perspective demonstrates the percentile bucket of $\hat{H}$. With higher resolution, the curvature shows a smoother transition from low to high percentiles.}\label{high-res}
\end{figure*}

Adjustment for inflation with the consumer price index $C^{0}_{t}$ is also crucial for proper income calculation. At period $t$, $C^{0}_{t}$ is the ratio of the market basket price $m_t$ to the base period $m_0$:
\begin{equation}
    C^{0}_{t} = \frac{m_t}{m_0}
    \label{cpi1}
\end{equation}
Assuming that the market basket structure is stable over time, we use the consumer price index as an approximate control for the price variation. This assignment allows us to modify household income $H$ for inflation. Although the index suffers from the biases (e.g.\ new good, quality, outlet, and substitution) that cause the overstate inflation, we know no other alternatives to control the $R$ changes. We also convert the base period of the consumer price index from 1999 to 2019. Therefore, the data is plausible to compare the income in the various time frames:
\begin{equation}
    C_t^{2019} = \frac{m_t}{m_{2019}} = \frac{m_t/m_{1999}}{m_{2019}/m_{1999}}= \frac{C_t^{1999}}{C_{2019}^{1999}} = \frac{C_t^{1999}}{0.652}
    \label{cpi2}
\end{equation}
The adequate household size $S$ should also be taken into account, since it affects the family income severely. For example, a household comprised of two people with total revenue of \$100,000 is more affordable than a six-person family with the same income. We avoid dividing $H$ by the total number of household members (i.e.\ $S^2$) since some expenses do not scale up linearly. For example, a two-bedroom apartment is not twice as expensive as the rent of a one-bedroom apartment. We decide to adjust for household size using the square root equivalent scale for consistency~\cite[page 13]{johnson2005}.

In theory, several other normalizers (e.g.\ taxes and transfers) should be added to adjust $H$ for more accurate comparisons. Unfortunately, the data for these normalizers are not available at the household level. Therefore, using $C, R, S$, we adjust the household income $\hat{H}$ as:
\begin{equation}
    \hat{H} = \frac{H}{CRS}
    \label{normalizer}
\end{equation}

\subsection{Age Adjustment}
Since age correlates strongly with income, it is not appropriate to compare $\hat{H}$ distribution with different age distribution~\cite{almas2012, formby1980, murphy1990}. Indeed, the population in CA ages faster than in DC. In 1976 both CA and DC age peak around twenty-five. As time progresses in Figure~\ref{age-6}, the age peak moves towards fifty. Meanwhile, Figure~\ref{age-11} shows that the curve's summit is around twenty-five consistently until 2019. These two states illustrate that different regions in the U.S. have various age distribution.

Therefore, comparing states' income is meaningless due to the relationship between age and income. To avoid this sampling issue, we proposed a sampling technique that standardized the age distribution. By transforming the data more uniformly distributed in each region, finding the similarities and differences between each state income is more plausible.

\subsection{Benchmark design}
The first approach of designing the x-axis is to sort the U.S. regions' median values. As Figure~\ref{ranking} suggested, this technique introduces an artificial amount of variance in the hierarchy of each state and imposes the fluctuation on the ranking system. Therefore, the relative ranking system does not necessarily correlate with real economic growth. Using this method, the ranking on the x-axis is not reliable for analysis.

To avoid these issues, we design a second method: the distance to 2019 $\hat{H}$ benchmark for our analysis. Each state's position on the x-axis is the difference between the fiftieth percentile of each year and 2019. This approach reduces the variation in the ranking system and visualizes the development rate of the economy in the U.S. more realistic.

\begin{figure*}[t]\centering
    \includegraphics[width= 0.32\textwidth]{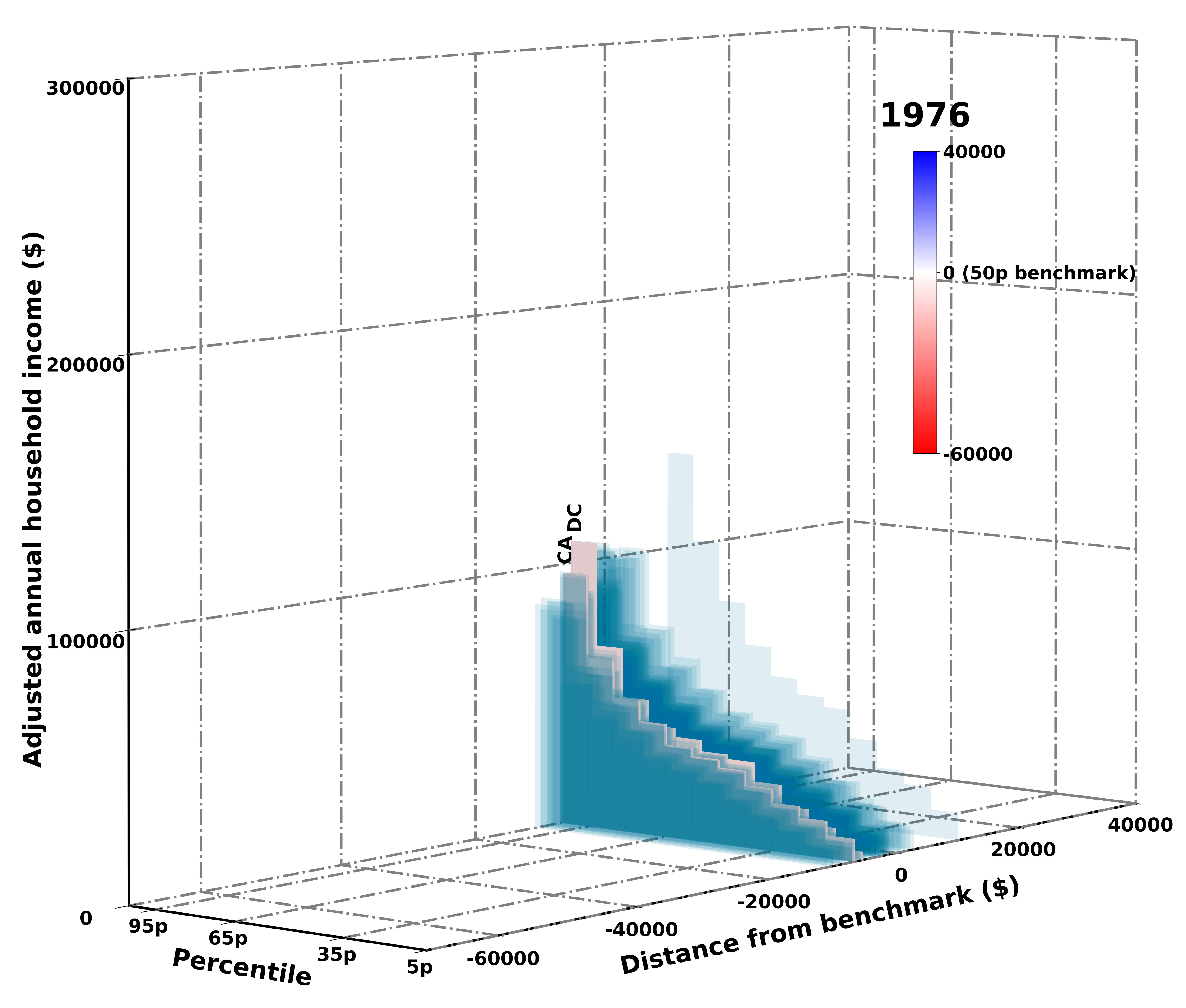}
    \includegraphics[width= 0.32\textwidth]{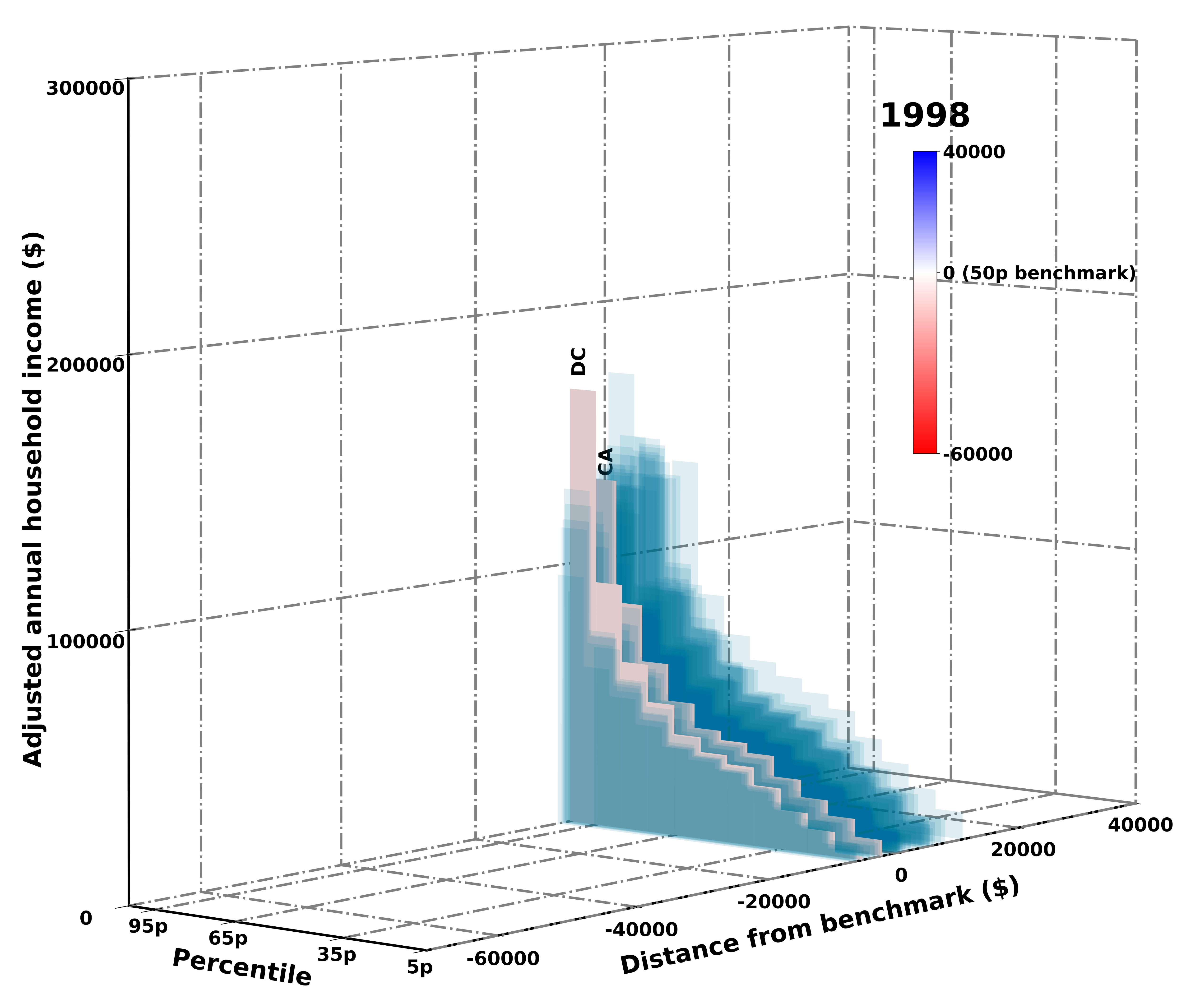}
    \includegraphics[width= 0.32\textwidth]{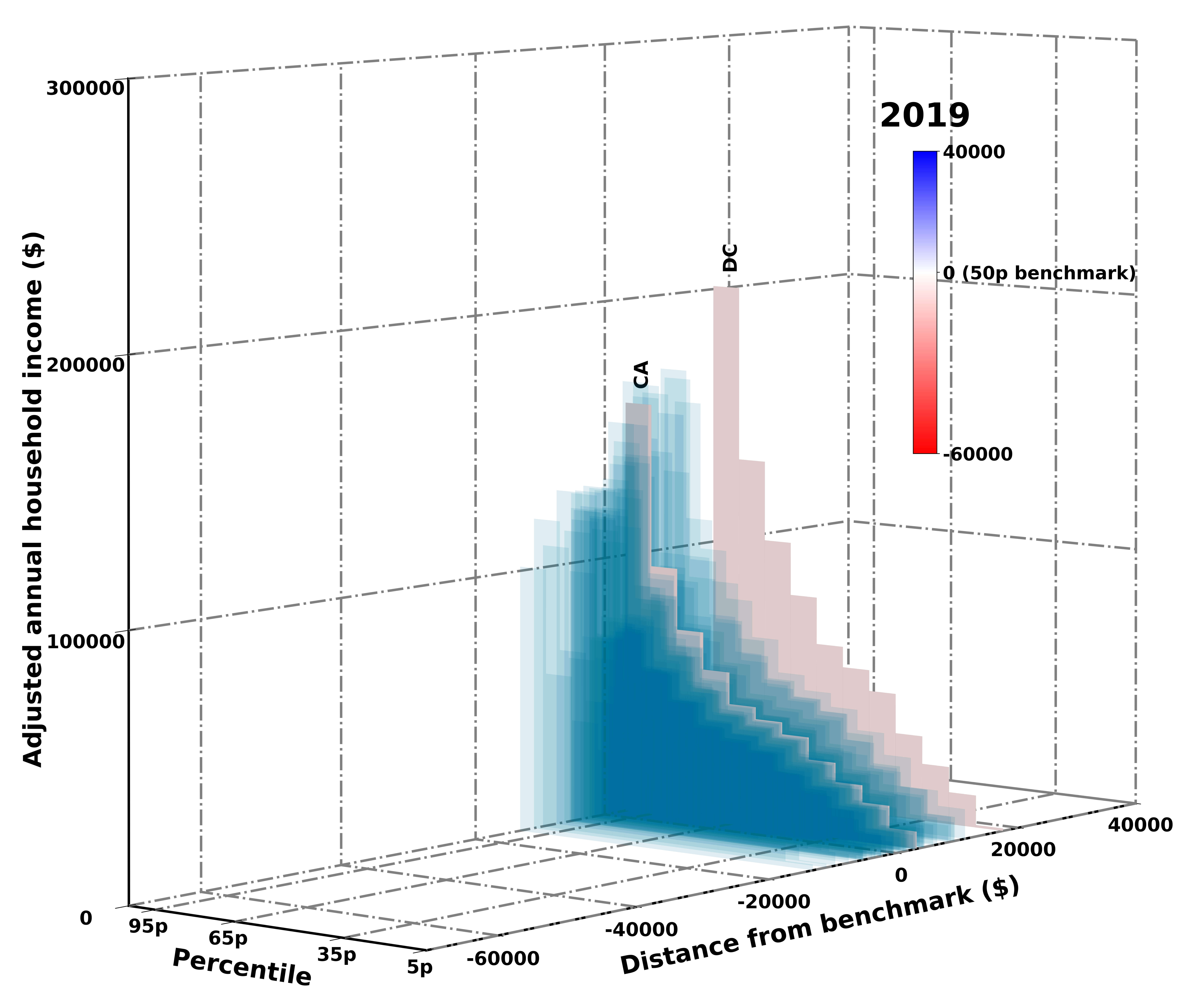}
    \caption{The fourth perspective highlights $\hat{H}$ of CA and DC from 1976 to 2019.}\label{state-highlight}
\end{figure*}
\begin{figure*}[t]\centering
    \includegraphics[width= 0.32\textwidth]{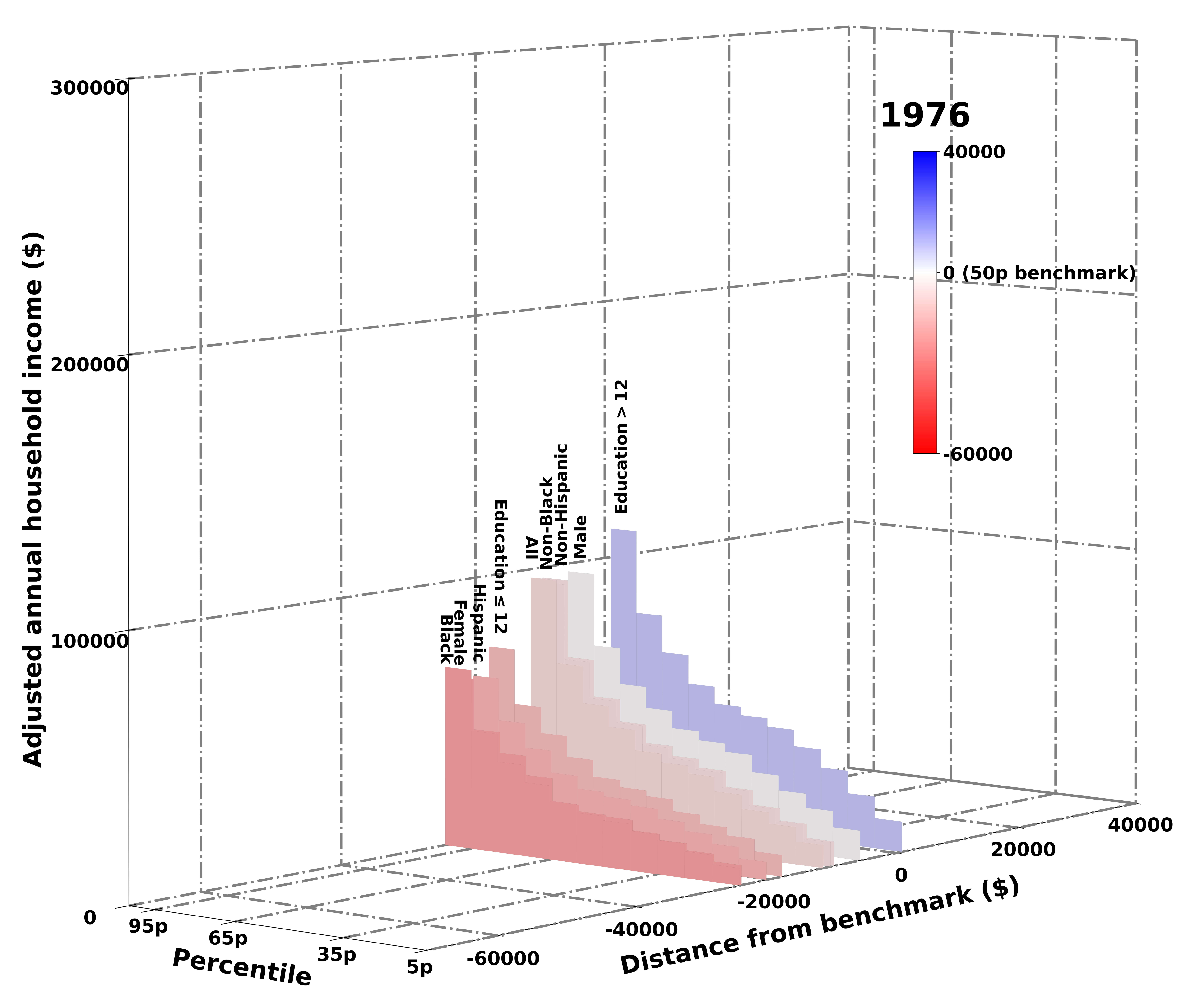}
    \includegraphics[width= 0.32\textwidth]{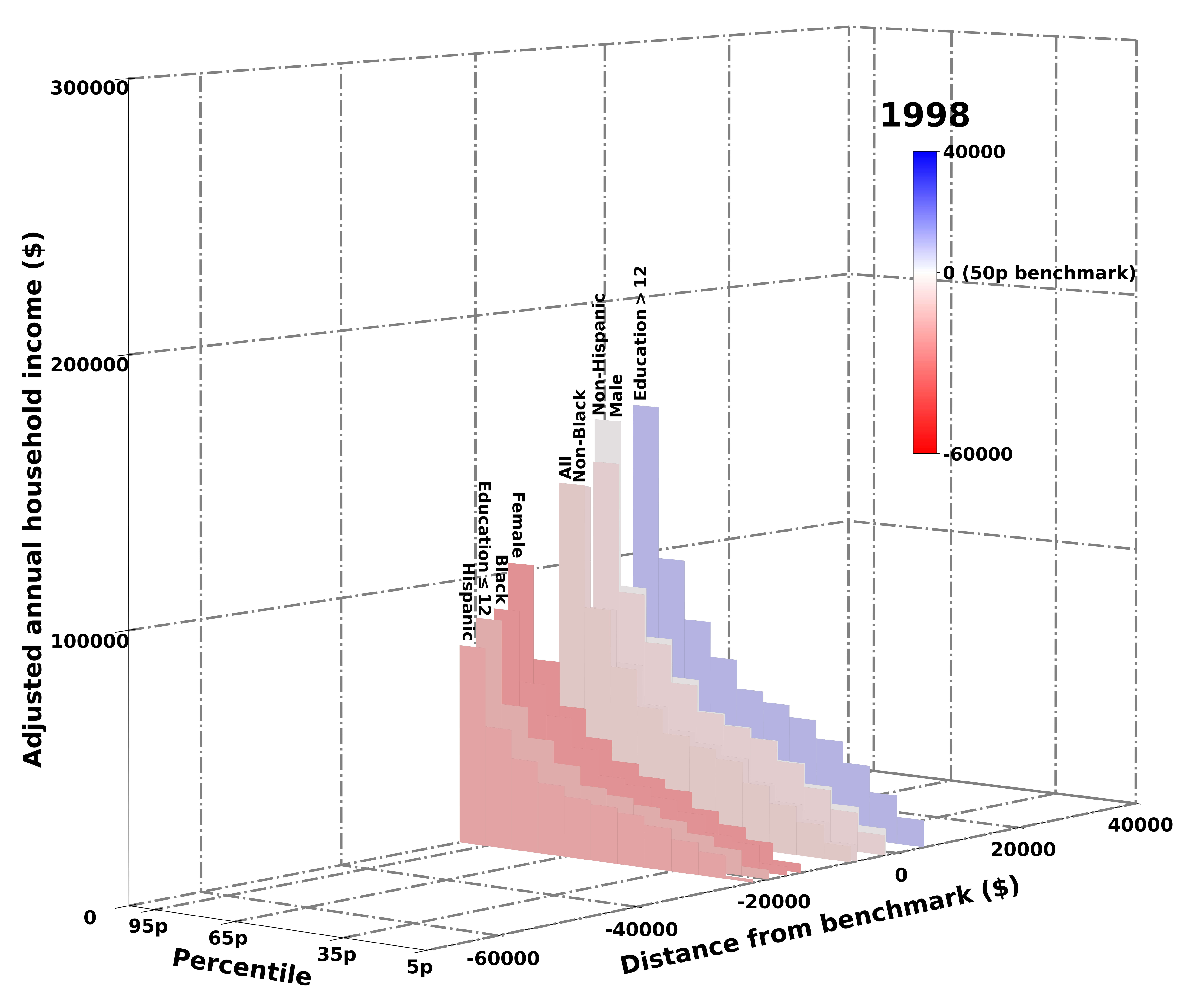}
    \includegraphics[width= 0.32\textwidth]{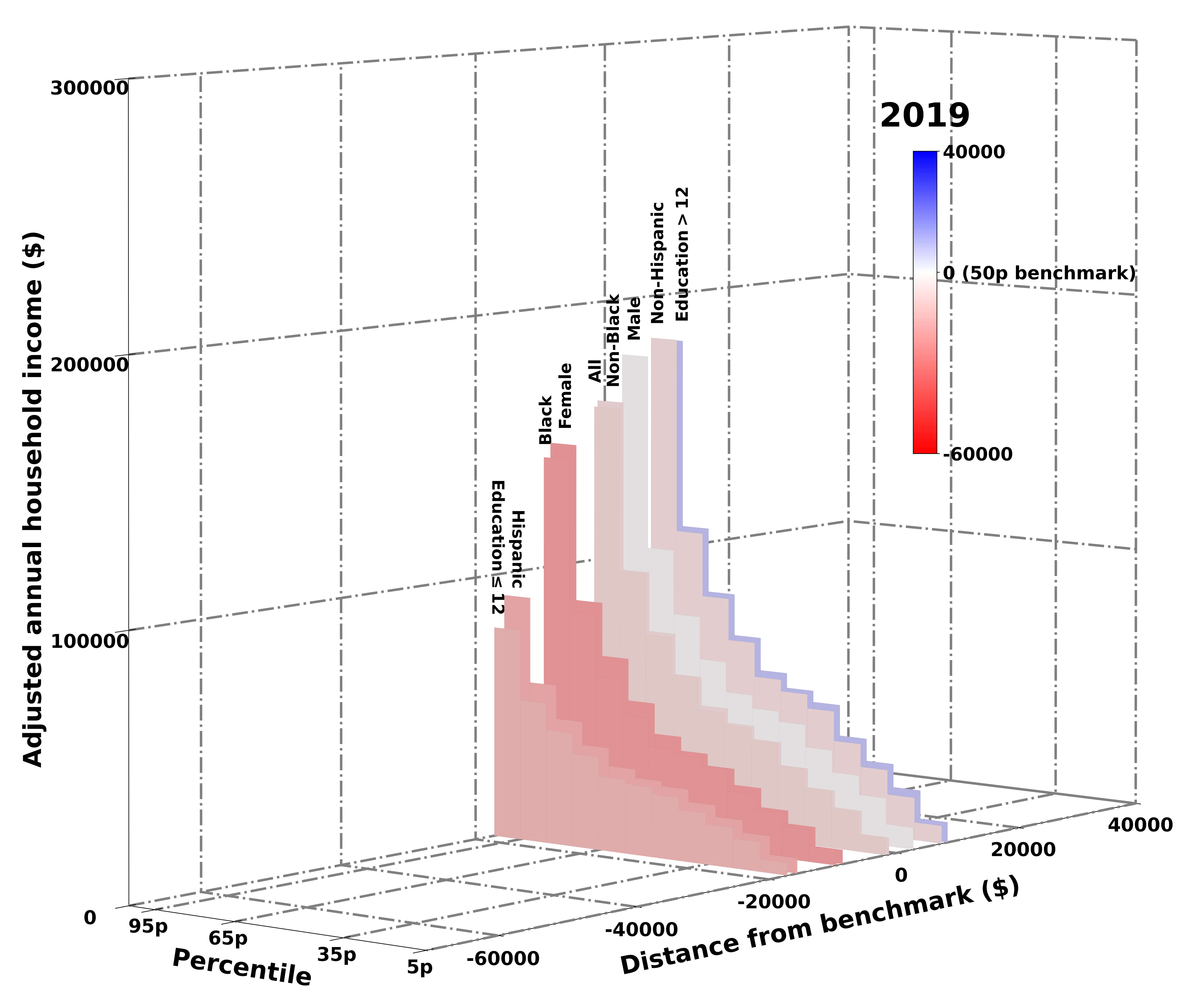}
    \caption{The fifth perspective is the adjusted household income with all and various subcategories in CA: Black, Non-Black, Hispanic, Non-Hispanic, Education with or without high school diploma (Education $\leq$ 12), Education with higher degrees after high school (Education $>$ 12), Male, and Female.}\label{subcat-highlight}
\end{figure*}

\subsection{Income Segmentation}
After the age distribution is standardized, $\hat{H}$ is segmented into percentile bucket $B_k$ from five to ninety-five. Household income can be negative because it includes various income sources, such as business income. Hence, we exclude the bottom five percentiles to avoid the negative value of household income. We also remove the top five percentiles to avoid invalid high-income measurement due to the disclosure avoidance measure of IPUMS-CPS survey~\cite{flood2020}. We then calculate the cumulative household weight ($w$) and the household percentile ($P$). For household $h$ in-state $s$ with total household $t$ at year $y$, $P$ is expressed as:
\begin{equation}
    P_{h, s, y} = \frac{\sum_{i=0}^{h} w_{i, s, y}}{\sum_{i=0}^{t} w_{i, s, y}}
    \label{percentile}
\end{equation}
Then, the adjusted household income is sorted in ascending percentile bucket. In this study, we also employ the decile bucket:
\begin{equation}
    \begin{split}
        \text{decile: } &k \in \{05\%, 15\%, ..., 45\%, 50\%, 55\%, ..., 95\%\}\\
        \text{percentile: } &k \in \{05\%, 06\%, ..., 49\%, 50\%, 51\%, ..., 95\%\}
    \end{split}
    \label{bucket_type}
\end{equation}
Given state $s$ and year $y$, a bucket $b_{k, s, y}$ is defined as:
\begin{equation}
    b_{k, s, y} = \{H_{h, s, y} | k-1 \leq P_{h, s, y} \leq k \}
    \label{bucket}
\end{equation}
We choose the maximum of $b_{k, s, y}$ as its summary statistic and the decile bucket as our primary measurement.

\subsection{Population Size Adjustment}
For state $s$ in year $y$ with $h$ households, the thickness of its slide is proportional to its normalized population $\bar{p}$, which computes as:
\begin{equation}
    \bar{p}_s = \frac{\sum_{i=0}^{h} w_i}{min(p_s)}
    \label{pop}
\end{equation}
where $w_i$ represents the weight of household $i$ in the sample.

\section{Result and Discussion}
\subsection{Normalizers for Household Income}
Data about actual and backcasted regional price parity of some example states is shown in  Figure~\ref{rpp}. We observe a moderate growth in regional price parity in all states. The value of fixed-effect coefficients of each state is presented in Figure~\ref{fe-coef}. CA, DC, and NY are the states with the highest geographic with the specific cost of living, even after controlling the gross rent.

\begin{figure*}[t] \centering
    \includegraphics[width= 0.32\textwidth]{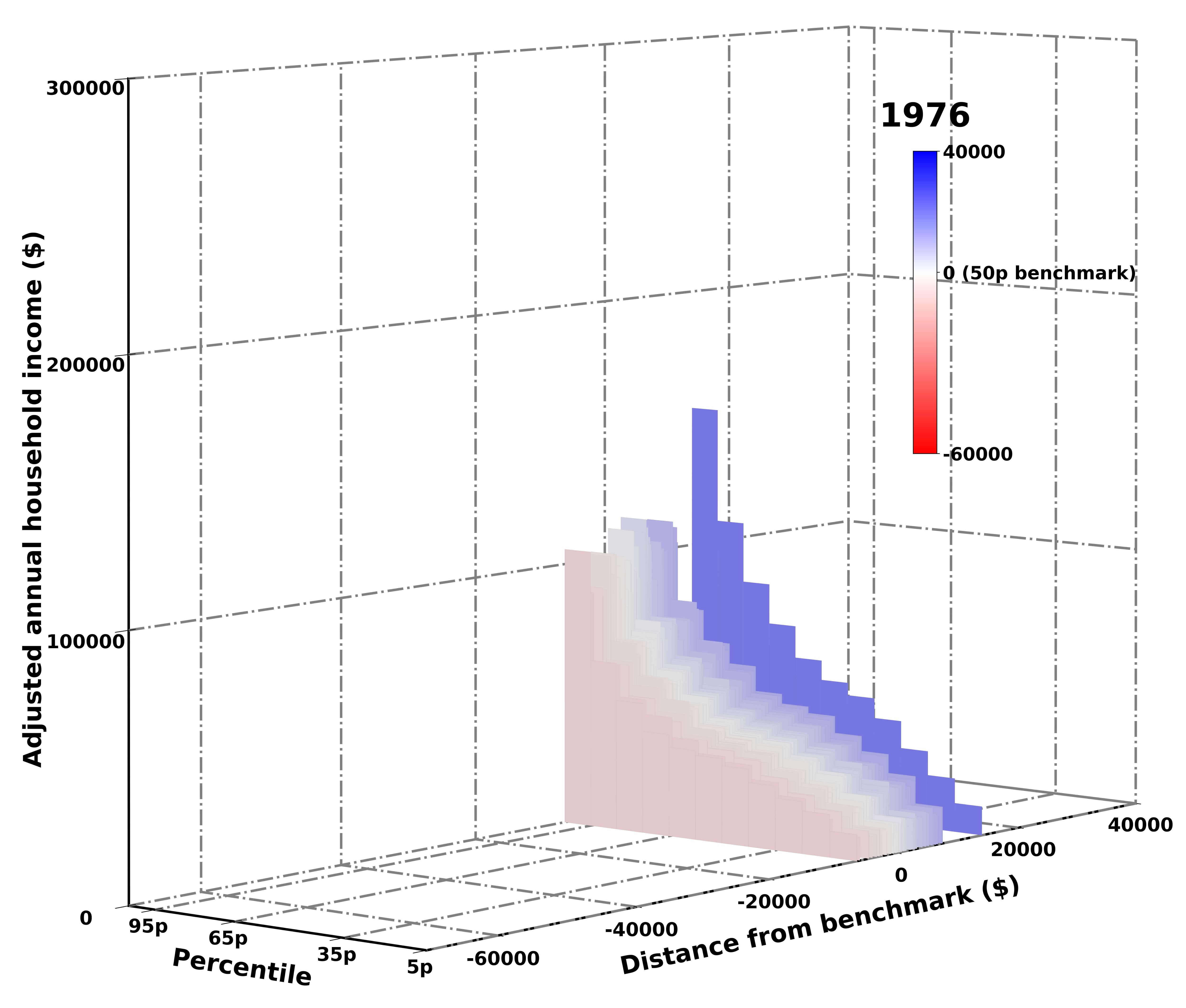}
    \includegraphics[width= 0.32\textwidth]{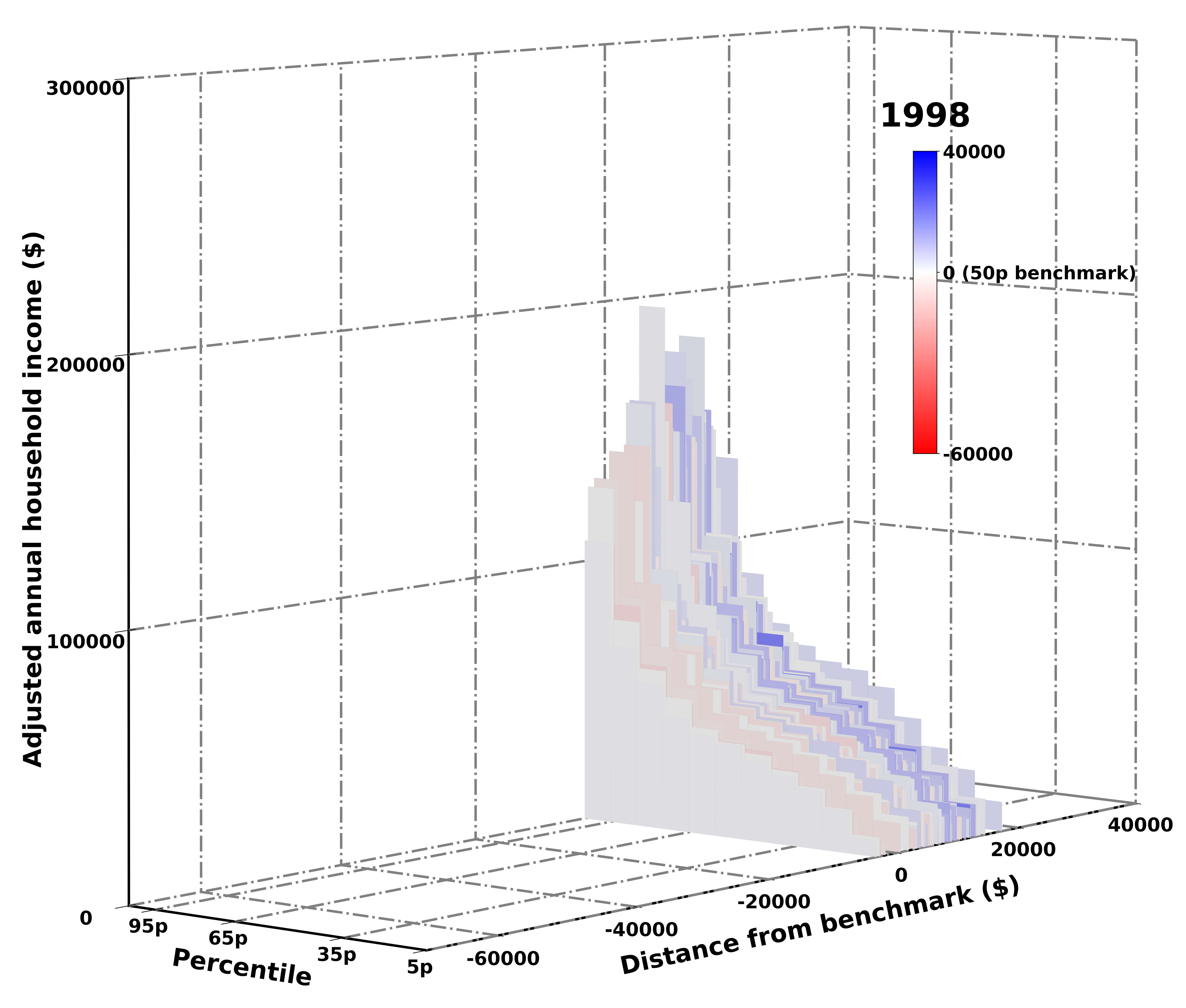}
    \includegraphics[width= 0.32\textwidth]{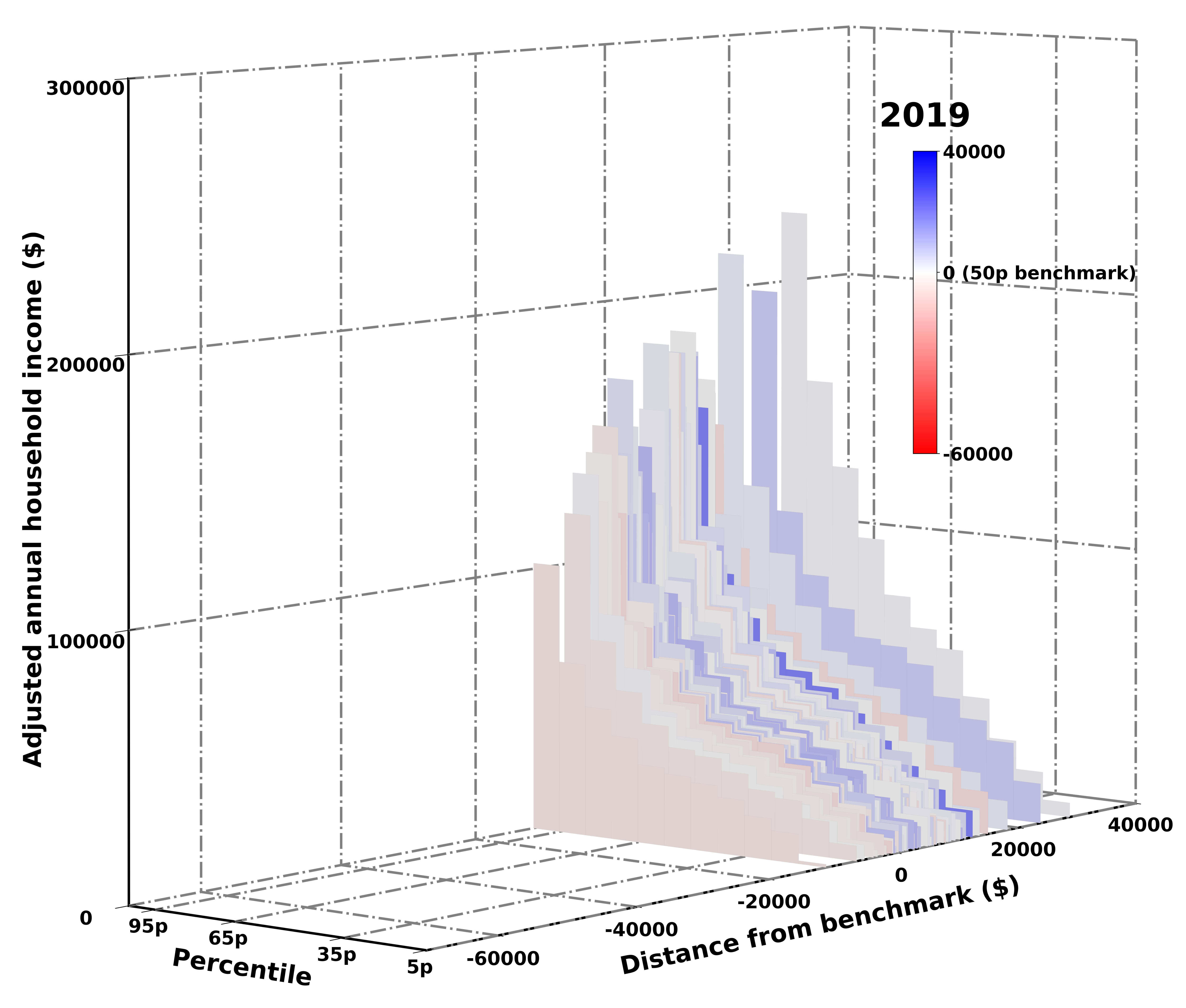}
    \caption{The sixth perspective shows $\hat{H}$ distribution of males only from 1976 to 2021.}\label{subpop_male}
\end{figure*}
\begin{figure*}[t] \centering
    \includegraphics[width= 0.32\textwidth]{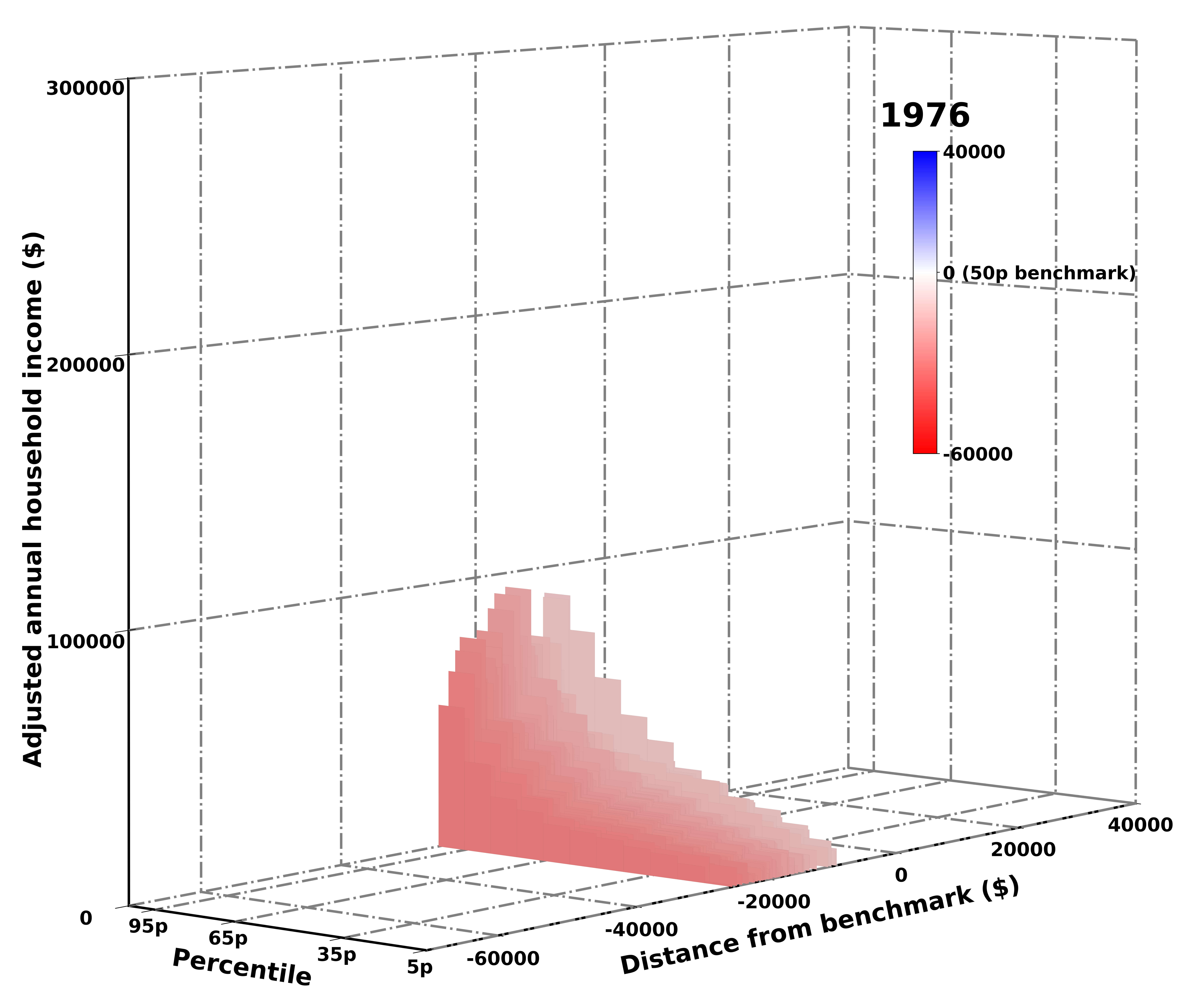}
    \includegraphics[width= 0.32\textwidth]{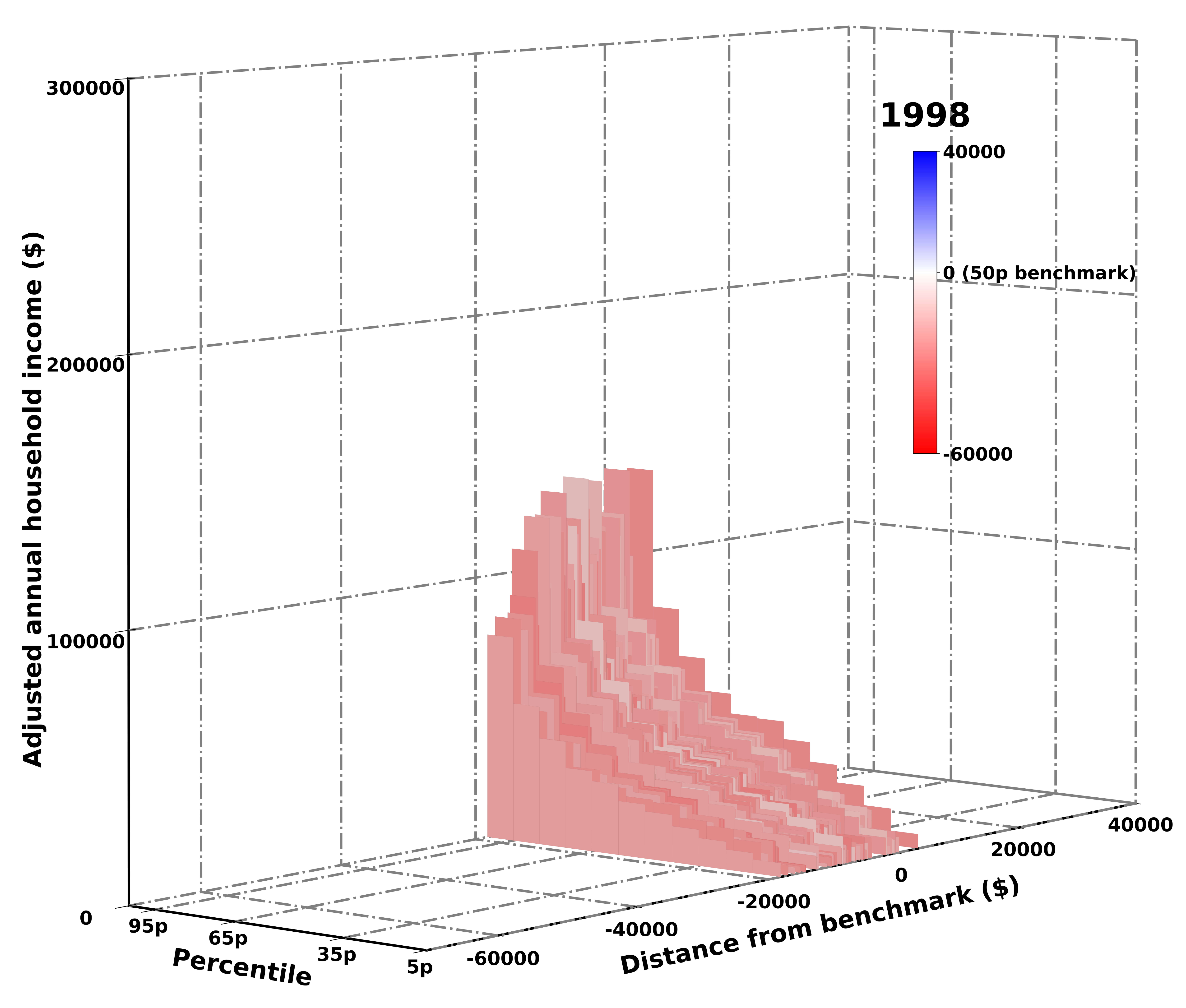}
    \includegraphics[width= 0.32\textwidth]{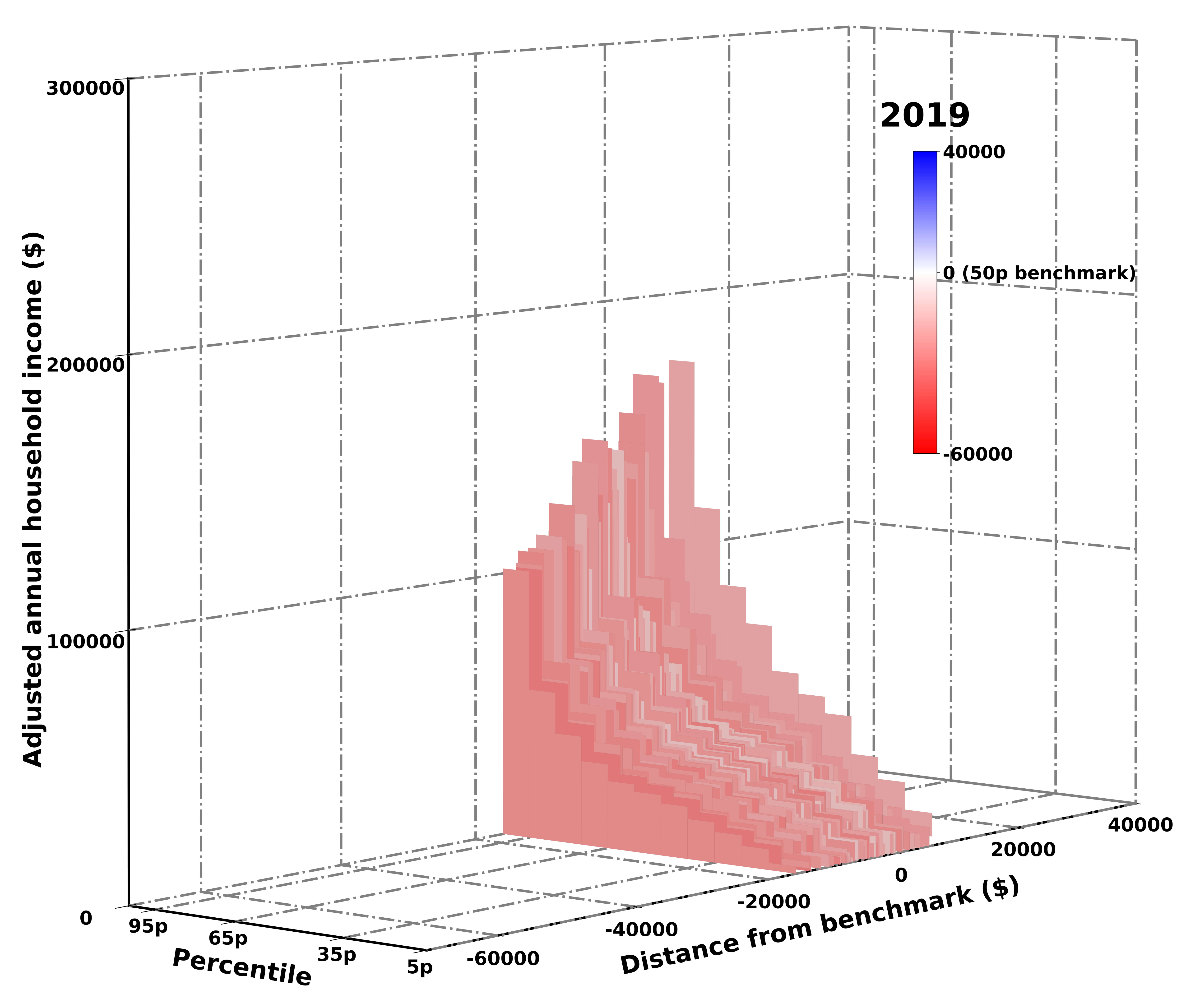}
    \caption{The seventh perspective shows $\hat{H}$ distribution of females only from 1976 to 2021.}\label{subpop_female}
\end{figure*}

\subsection{Seven Perspectives in U.S. Income Distribution}
Figure~\ref{h} demonstrates the development of unadjusted household income distribution $H$. Unlike other perspectives, the growth rate of $H$ is much faster than $\hat{H}$. In 1976, all the benchmarks varied from -55000 to -40000. In 2019, all states ended up from -20000 to 25000. Even though the economic growth rate is magnificient, this development leads to an illusion. In Figure~\ref{h}, the U.S. economy advances remarkably due to the shift of unadjusted income. However, in Figure~\ref{hhat}, the economy overall does not progress much since $\hat{H}$ does not shift much as $h$. Hence, the income grows in accommodation to other payments and the family size. Figure~\ref{rpp-backcast} illustrates the expansion of cost of living and household size. Therefore, the economy does not grow significantly, as figure~\ref{h} suggested.

Moreover, visualizing the dynamic change between states in the U.S. is nearly impossible. Since in figure~\ref{h} within 1976, every state has quite similar benchmarks. Therefore, as time flies, even though the states' rank changes, the movement is not easily visible since their color is quite identical. Since Figure~\ref{hhat} emits the dynamic change among states and creates an illusion of growth, we decide the adjusted annual household income as our primary illustration for the rest of the perspectives.

Therefore, we adjust the income for Figure~\ref{hhat} which demonstrates the inequality expansion of $\hat{H}$ distribution in the U.S. from 1976 to 2019 more precisely. In 1976, each state was roughly linear (until the back wall). In 2019, the curvature from the front ($5^{th}$ percentile) to the back ($95^{th}$ percentile) is more abundant since the back blocks are much taller. Indeed, within the same year in Figure~\ref{hhat}, the $95^{th}$ percentile in the wealthier states is over \$150,000, while D.C., the thin slice with the tallest block, is around \$200,000. Besides, the front blocks (the poorest) stays short nearly the same from 1976 to 2019. With the increasing curvature, the gap between the lowest and highest household income enlarges over time.

Moreover, in the second perspective, the states' dynamic movement is much more pronounced. Assigning 1976 as the base year, the benchmark of each state changes over time. For example, in Figure~\ref{hhat}, in 1976 and 1998, D.C. blends into different states since their colors are roughly the same: pink and white. However, in 2019, D.C. stands out from the others and attains the benchmark of around 20000. Not only D.C. but also various states alter their positions as time progresses.

The third perspective demonstrates a higher resolution of $\hat{H}$. Compared to the Figure~\ref{hhat}, the curvature in Figure~\ref{high-res} is smoother since it applies the percentile bucket. Therefore, it demonstrates a more precise linearity scale of percentile income inequality in each state. However, with the decile bucket, the income inequality is still visible. Therefore, to remove the unimportant percentiles, we primarily deploy the decile bucket for other perspectives.

\begin{figure}[hbt!]\centering
    \includegraphics[width = \columnwidth]{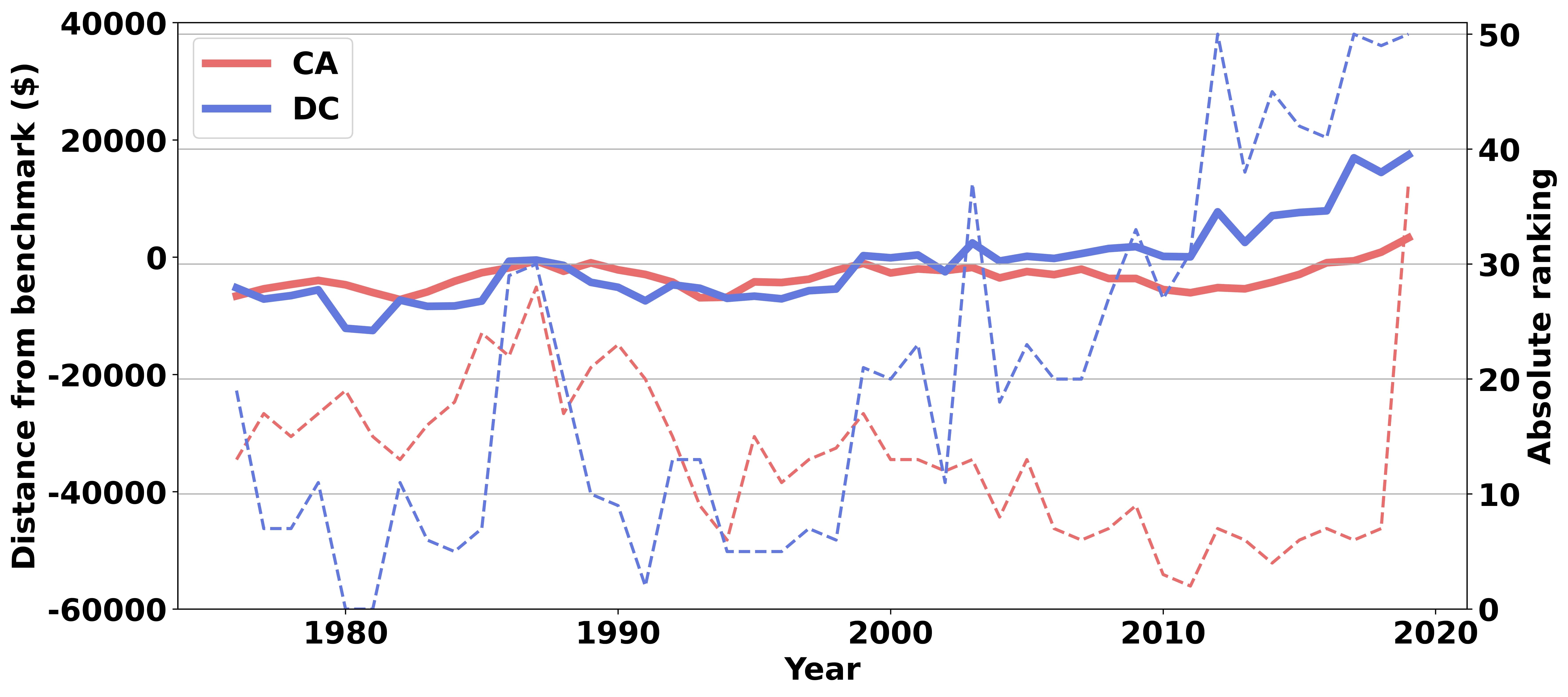}
    \caption{The relative ranking and distance to benchmark of CA and DC over time.}\label{ranking}
\end{figure}

The fourth perspective highlights C.A. and D.C. to visualize the dynamic movement changes more clearly. In 1976 in Figure~\ref{state-highlight}, since C.A. is on the left of D.C., it is generally poorer than D.C. This relative position between C.A. and D.C. is also the same in 2019, but with a more significant gap. However, in 1998, C.A. is wealthier than D.C. since it is on the right of D.C. More precisely, before the 2000s, C.A. and D.C.'s position is frequently close to each other, as Figure~\ref{ranking} suggested. However, after that period, C.A. and D.C.'s distance grows apart. Eventually, DC attains a higher benchmark and rank.

Although people in D.C. are generally wealthier than in C.A., they also suffer from more income inequality. As the percentile rises in Figure~\ref{sampling_var}, the peak of $\hat{H}$ in D.C. is more significant than in C.A. Moreover, the density of $\hat{H}$ in D.C. is more skewed to the right than in C.A. Therefore, although D.C. people generally earn more than C.A. people, its income inequality is more severe.

Since adjusted income distribution in C.A. is close to the normal distribution, we investigate some demographic aspects correlated with the income. From 1976 to 2019 in Figure~\ref{subcat-highlight}, Male, Education higher than the high school level, and people who are neither Black nor Hispanic are more dominant at $\hat{H}$. Indeed, in 1976, females, Black, or Hispanic make even less than people who have education less than a high school diploma. These groups had low income before 2020, but as time progress, their position change. In 2019, females, Black, and Hispanic move closer to C.A. with all demographic aspects.


The sixth and seventh perspective demonstrates more explicitly the decline in income inequality among genders. In 1976, the dominant color in figure~\ref{subpop_male} was blue, and most of the states were more than zero. On the other hand, within the same year, the dominant color in figure~\ref{subpop_female} was red, and every state had a benchmark less than zero. These two dominant colors suggest that in 1976, the female does not earn as high as the male. However, as the years pass, women have more chances of making a higher income. In 2019, Figure~\ref{subpop_female} demonstrates that most states have females' benchmarks larger than 0. Therefore, the income inequality in genders is reducing eventually. However, even now, men still earn more than women in general.

\subsection{Sampling Variability}
\begin{figure}[hbt!]\centering
    \includegraphics[width = \columnwidth]{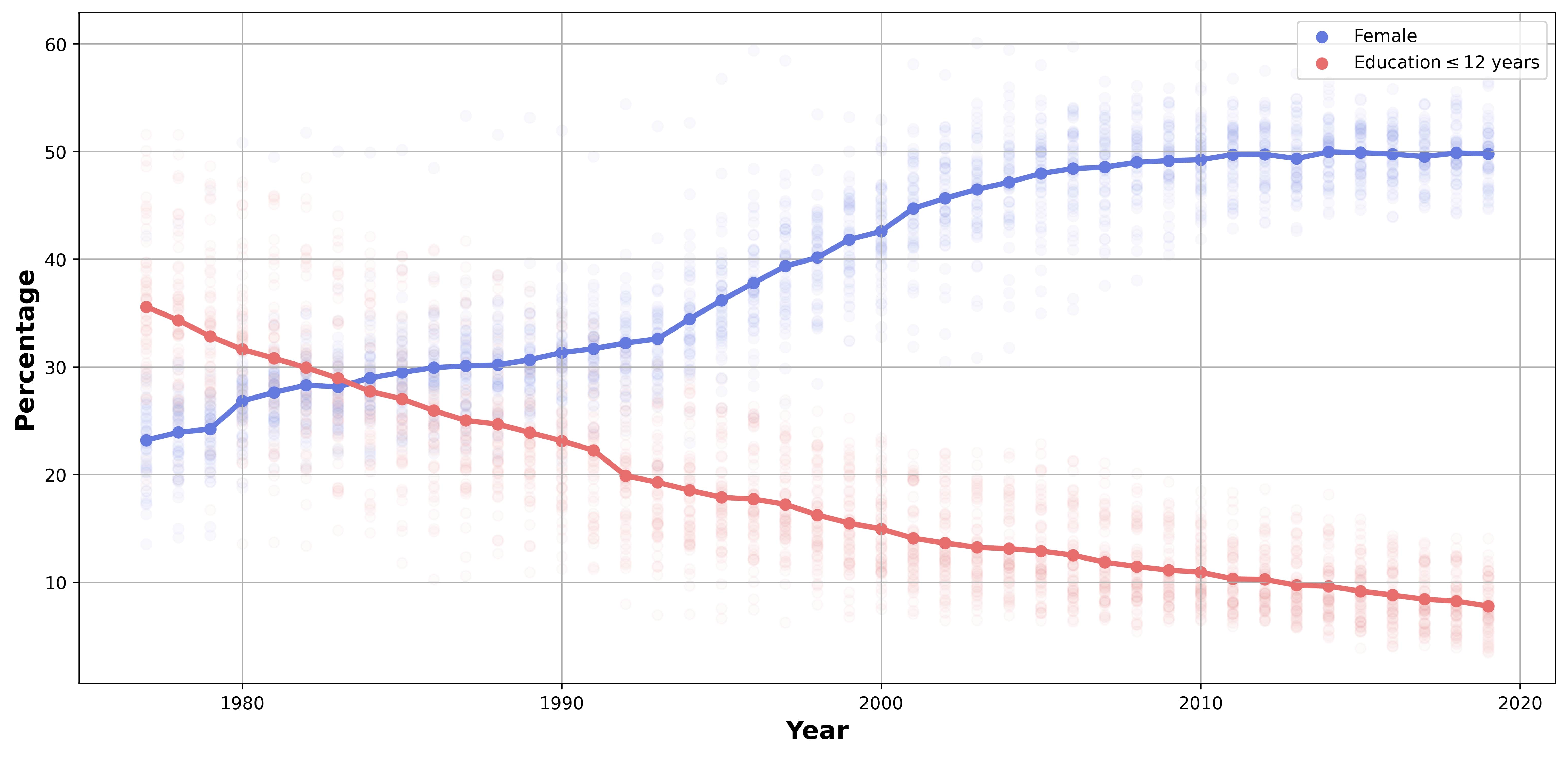}
    \caption{The percentage of households with high school or below education are declined drastically over time. Conversely, the percentage of the female household approaches 50\%.}\label{datadist}
\end{figure}

The information communicated through our visualization best interprets with sampling variability due to its empirical nature. The height of each block in Figure~\ref{hhat} is a sample estimate of $\bar{H}$ percentile for a state. Therefore, it has a standard error (S.E.). To estimate S.E.'s magnitude for each block's height depicted in the charts, we computed the bootstrapped S.E.'s for D.C. and CA in 1976 and 2019. We choose to demonstrate the sampling variable with C.A. and D.C. because they are consistent with one region, which has the most prominent and smallest sample size, respectively.

\begin{figure}[t]\centering
    \includegraphics[width = \columnwidth]{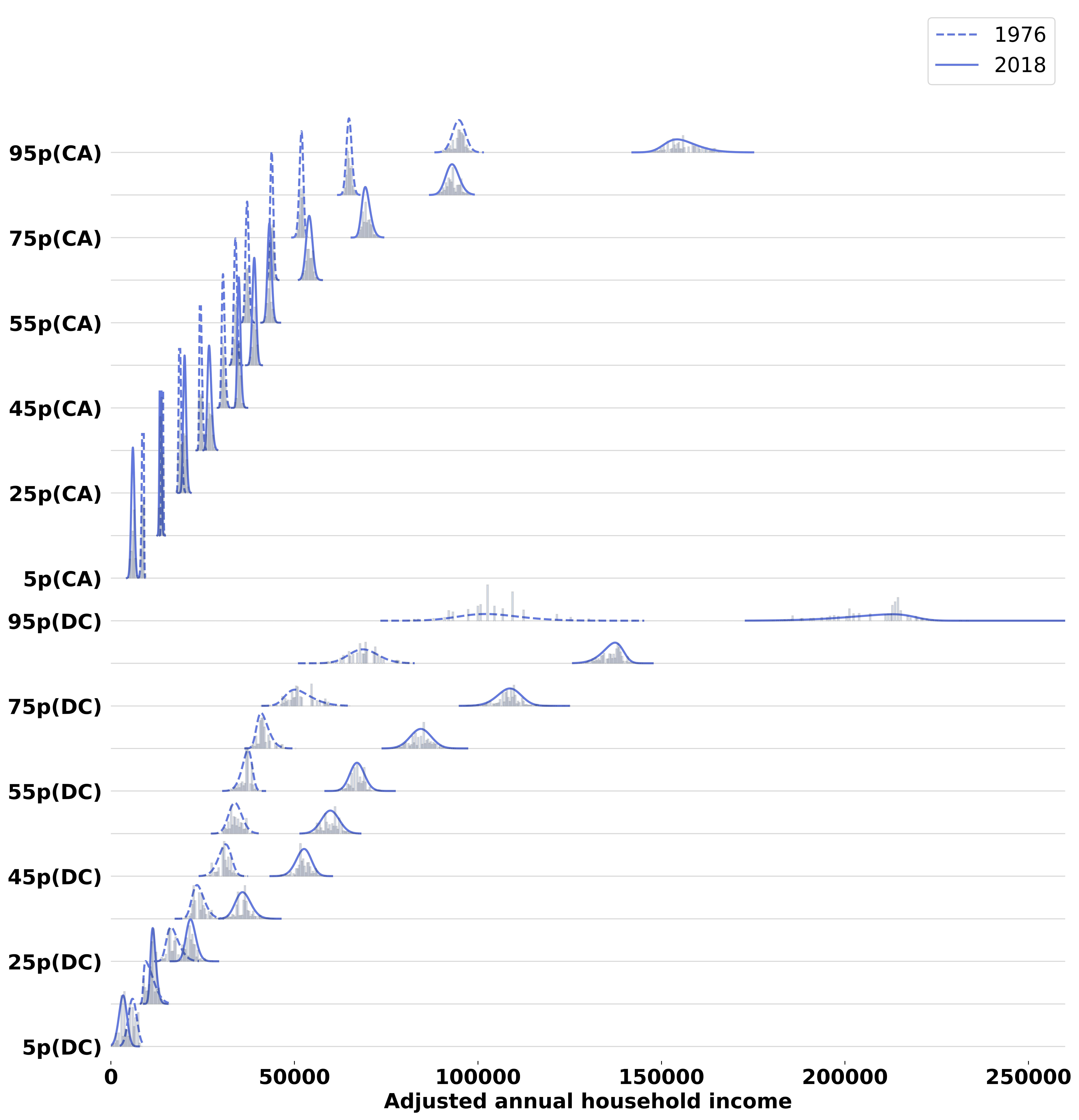}
    \caption{Adjusted annual household income distribution in CA and DC in 1976 and 2019.}\label{sampling_var}
\end{figure}

In Figure~\ref{sampling_var}, the dollar value of the standard error for percentile estimation rises as the percentile rises and falls as the sample size increases. The S.E. of highly populated states is several thousand dollars for the $95^{th}$ percentile, but for states with thin slices (e.g.\ D.C. and S.D.), the bounce is in the tens of thousands of dollars. For example, in Figure~\ref{sampling_var}, the margin of error in C.A. is generally smaller than in D.C. since C.A. has a higher population. In other words, if the Bureau of Labor Statistics had carried out a second survey for 1976, we would see the $95^{th}$ percentile of real household income in D.C. vary by plus or minus roughly \$13,500.

We emphasize that there is sampling variability in the chart. We should interpret the blocks' heights as estimations. Moreover, these heights are prone to sampling fluctuations that increase as we go from front to back. A sharp-eyed viewer may notice that in Figure~\ref{hhat}, the back wall in 2019 is not only taller but more jagged than it was in 1976. This observation cannot naively interpret the back block based on the actual population $95^{th}$ percentile\textemdash sampling variability must be part of the discussion.

It is easy to see that most of D.C.'s back wall fluctuations are too large to be explained by real economic forces\textemdash. These ups and downs reflect the effects of random sampling. C.A.'s back wall is smoother than D.C.'s because its sample size is much more significant, which is 5412 versus 320 observations.

\subsection{Interactive Visualization}
Parallel to the python package, we also implement our visualization using JavaScript for further engagement with the users. This interactive visualization is executable on any website. A demonstration of our interactive version is at https://github.com/sangttruong/incomevis. The users can hover their mouse over different parts of the chart to get a popup with information. The down arrow in the screen's top-right corner allows the users to download the chart's data for other income computation. They can also apply individual state labels: choose the year; and display either real household income (RHH) or real household income per (equivalized) person (ERHH). The latter demonstrates the same four points as above but controls the household size, giving a better measure of household income per person. The online version of the chart invites comparison and generates questions.

Moreover, our website provides an option for selecting real household income adjusted for state prices (RHHRPP) and real household income per (equivalized) person modified for state prices (ERHHRPP) from 2008 to 2019. For teaching or exploration purpose, customization of the graph is available on our interactive version. Also, the animation function is available with playback controls at https://github.com/sangttruong/incomevis.

\section{Conclusion and Further Research}
We introduce a framework for visualizing income distribution in the U.S. Regional price parity, consumer price index, and household size are the adjustment variables for our data. Therefore, our paper can demonstrate a robust comparison of income. Also, to control different demographic and age distribution, we apply the resampling technique.  Moreover, our visualization communicates more explicitly to the public audience than the Gini coefficient because of no background technical knowledge requirement.

As with any other empirical long-term population study, our analysis identifies some underlying limitations. For example, the CPS ASEC survey is not the same every year. Questions change, and so do data collection methods (e.g., see a discussion of majority changes in 2017 at~\cite{rothbaum2019}). Nonetheless, we provide an entr{\'e}e to the income distribution and inequality study. A visualization for household income in the U.S. over time offers an excellent starting point because it lays bare the facts, captures attention, and stimulates many questions about causes and remedies.

\section{Acknowledgement}
This project was started by Sang Truong and Humberto Barreto during summer 2019 under the support of Hewlett Mellon Presidential Fund for undergraduate research at DePauw University. We thank Frank Howland, Jonah Barreto, Jarod Hunt, and Bu Tran their support on preparing the manuscript.

\bibliographystyle{abbrv-doi}
\bibliography{main}
\end{document}